\documentclass[11pt,english]{article}
\usepackage[T1]{fontenc}
\usepackage[latin9]{inputenc}
\usepackage{amsmath}
\usepackage{amssymb}
\usepackage{graphicx}
\usepackage[numbers]{natbib}

\makeatletter
\pdfoutput=1
\usepackage[T1]{fontenc}
\usepackage[latin9]{inputenc}
\usepackage{color}
\usepackage{float}
\usepackage{graphicx}
\usepackage{graphics}
\usepackage{esint}
\usepackage{enumerate}

\makeatletter

\usepackage{latexsym}
\usepackage{epsfig}
\usepackage{graphicx}
\usepackage[colorlinks,linkcolor=blue]{hyperref}
\allowdisplaybreaks 

\usepackage{babel}
\usepackage{tcolorbox}

\usepackage{tikz}

\makeatother

\usepackage{babel}
\begin{document}
\vspace*{10.0ex}
\centerline{\Large \bf Entanglement entropy from fundamental strings}

\vspace*{10.0ex}
\centerline{\large  Houwen Wu and Shuxuan Ying}
\vspace*{7.0ex}
\vspace*{2.0ex}

\centerline{\large \it College of Physics}
\centerline{\large \it Sichuan University}
\centerline{\large \it Chengdu, 610065, China} \vspace*{1.0ex}

\vspace*{3.0ex}
\centerline{\large \it Department of Physics}
\centerline{\large \it Chongqing University}
\centerline{\large \it Chongqing, 401331, China} \vspace*{1.0ex}
\vspace*{3.0ex}

\centerline{iverwu@scu.edu.cn, ysxuan@cqu.edu.cn}
\vspace*{8.0ex}
\centerline{\bf Abstract} \bigskip \smallskip
In previous work, we derived the local entropy capacity from off-shell string theory and combined it with the gravitational replica construction and flow-cut duality to normalize the entropy flow carried by physical fundamental strings. In this paper, we apply this framework to interval entanglement in holographic CFT$_2$. We construct explicit on-shell F1 worldsheets in the AdS$_3$ vacuum and planar BTZ, derive their Kalb-Ramond source currents, and show that the normalized spatial currents saturate the corresponding RT surface. Their flux reproduces both the vacuum single interval entropy and the finite temperature CFT$_2$ result. We then vary the NS-NS flux fraction. Although the total maximal flux remains unchanged, the F1 equations continuously redistribute the strings among different parts of the complement. In planar BTZ, the resulting signed flux through the bifurcation line reproduces the extensive thermal entropy and can vanish in the pure NS-NS limit without changing the interval entropy. Thus the on-shell string dynamics additionally determines its physical routing and boundary contact structure. 
\vfill 
\eject
\baselineskip=16pt
\vspace*{10.0ex}
\tableofcontents

\section{Introduction}

\label{sec:introduction}

The Ryu--Takayanagi (RT) formula translates the entanglement entropy
of a boundary region into the area of a bulk minimal surface \cite{Ryu:2006bv,Ryu:2006ef}.
On a static bulk slice, the bit-thread formulation replaces this minimal
surface by a maximal flow \cite{Freedman:2016zud,Headrick:2017ucz},
\begin{equation}
S\left(A\right)=\max_{v}\int_{A}v^{i}n_{i}\,dA_{E},\qquad\nabla_{i}^{E}v^{i}=0,\qquad|v|_{E}\leq\frac{1}{4G_{N}},\label{eq:intro bitthreads}
\end{equation}
where the metric, norm, and area element in this expression are in
the Einstein frame. Conservation makes the flux invariant under deformations
among homologous cuts, while the norm bound limits the entropy flux
through each unit area. The RT surface is the bottleneck at which
the maximal flow saturates this bound. This formulation has led to
geometric constructions of holographic flows \cite{Agon:2018lwq}
and direct proofs of entropy inequalities \cite{Freedman:2016zud,Hubeny:2018bri},
as well as extensions to thermal states \cite{Mintchev:2022fcp,Caggioli:2024uza}.

The flow $v^{i}$, however, is normally introduced as the dual variable
of a convex optimization problem. Its integral curves may be represented
as threads, but the variational problem by itself does not identify
a spacetime charge or a physical object whose equations of motion
select those curves. A stringy realization should therefore answer
two separate questions. First, what fixes the local entropy capacity
that appears in the norm bound? Second, can a current carried by physical
strings realize a maximal flow and reproduce entanglement entropy?

Our previous work identified the spatial projection of the Kalb--Ramond
string charge density with an entropy flow and used it to reproduce
black hole entropies in three- and ten-dimensional backgrounds \cite{Wu:2025qwc,Wu:2026qha,Wu:2026kci}.
For certain extremal black holes, Tseytlin \cite{Tseytlin:1996qg}
obtained the entropy microscopically by counting string oscillator
states in a conformal sigma model. A subsequent construction related
the string-frame capacity to the genus-zero off-shell closed string
action and exhibited stationary on-shell F1 flows for black hole entropy
\cite{Wu:2026pgc}. The present paper addresses the corresponding
interval problem. We construct explicit families of on-shell F1 worldsheets
in the AdS$_{3}$ vacuum and in planar BTZ, derive their spacetime
currents directly from the F1 source action, and show that their spatial
projections reproduce the vacuum and thermal entanglement entropies
of a single CFT$_{2}$ interval. The probe limit is used throughout:
the F1 embeddings solve their complete worldsheet equations in fixed
background fields, while their backreaction is neglected. The previous
string and worldsheet based analyses of entanglement and black hole
entropy include Refs. \cite{He:2014gva,Brustein:2022wiq,Ahmadain:2022tew,Ahmadain:2022eso,Halder:2023adw,Mori:2025qoh,Ahmadain:2025pox}.

We briefly recall the string-frame normalization obtained in \cite{Wu:2026pgc}.
At genus zero and two-derivative order, the regulated closed-string
sphere gives the Euclidean curvature term \cite{Ahmadain:2022tew,Ahmadain:2022eso}
\begin{equation}
I_{\mathrm{closed}}=-\frac{1}{16\pi G_{N}^{\left(3\right)}}\int d^{3}x\sqrt{g}e^{-2\phi}R\left[g\right].\label{eq:review closed curvature}
\end{equation}
For a replica cone of opening angle $2\pi n$ around a curve $m$,
the singular curvature gives \cite{Fursaev:1995ef} 
\begin{equation}
\left.\left(I_{n}-nI_{1}\right)\right|_{\mathrm{tip}}=\left(n-1\right)\int_{m}\rho_{S}dA_{S}+O\left(\left(n-1\right)^{2}\right),\qquad\rho_{S}=\frac{e^{-2\phi_{0}}}{4G_{N}^{\left(3\right)}},\label{eq:intro string capacity}
\end{equation}
for constant $\phi=\phi_{0}$. The smooth NS--NS matter terms do
not add a localized tip contribution at this order. The Brown--Henneaux
central charge, expressed in the string frame, is \cite{Brown:1986nw}
\begin{equation}
c=\frac{3\ell_{\mathrm{AdS}}}{2G_{\mathrm{eff}}^{\left(3\right)}}=6\rho_{S}\ell_{\mathrm{AdS}},\qquad G_{\mathrm{eff}}^{\left(3\right)}=G_{N}^{\left(3\right)}e^{2\phi_{0}}.\label{eq:review central charge}
\end{equation}
The replica construction and flow--cut duality then give \cite{Freedman:2016zud,Headrick:2017ucz,Wu:2026pgc,Lewkowycz:2013nqa}
\begin{equation}
S_{A}^{\left(0\right)}=\min_{m\sim A_{\mathrm{reg}}}\int_{m}\rho_{S}dA_{S}=\max_{v}\Phi_{A}\left[v\right],\qquad\nabla_{i}v^{i}=0,\qquad|v|\leq\rho_{S},\label{eq:intro string frame flow}
\end{equation}
where all geometric quantities in the last expression are defined
using the spatial string-frame metric. This closed string result fixes
the capacity. The on-shell F1 current fixes a physical direction for
the entropy flow.

Our first result concerns the vacuum interval $A=\left(-R,R\right)$
in Poincare AdS$_{3}$. We obtain a one-parameter family of F1 worldsheets
whose time-zero profiles obey 

\begin{equation}
\left(x-R\right)^{2}+\left(z-z_{0}\right)^{2}=z_{0}^{2}.
\end{equation}
The worldsheets satisfy both the embedding equations, including the
Kalb--Ramond force, and the worldsheet metric constraints. We select
an ensemble of strings whose spatial profiles at $t=0$ intersect
the regulated RT surface at equal intervals of proper length along
the surface. We then coarse-grain the discrete F1 current and project
it onto the $t=0$ slice. We normalize the projected string charge
density to have norm $\rho_{S}$ within the string bundle. The resulting
flow is conserved and equals $\rho_{S}$ times the oriented unit normal
along the entire regulated RT surface. It therefore attains the maximal
flux,

\begin{equation}
S_{A}^{\left(0\right)}=\Phi_{A}=2\rho_{S}\ell_{\mathrm{AdS}}\mathrm{arc}\cosh\frac{R}{\varepsilon}=\frac{c}{3}\log\frac{2R}{\varepsilon}+O\left(\frac{\varepsilon^{2}}{R^{2}}\right).\label{eq:intro vacuum result}
\end{equation}
This gives the standard vacuum interval entropy \cite{Holzhey:1994we,Calabrese:2004eu}
directly from physical F1 probes.

Classical string propagation in BTZ, including probes in the NS--NS
background, has been studied in Refs. \cite{Larsen:1994ah,Ashok:2021ffx}.
We next construct the corresponding worldsheets in planar BTZ for
an interval $A=\left(x_{1},x_{2}\right)$ of length $L_{A}$. The
full embedding, its current, and the string spacing are derived independently
in the thermal coordinates. On the $t_{B}=0$ slice, the normalized
F1 current again saturates the RT surface normally. Its regulated
flux is $2\rho_{S}\ell_{\mathrm{AdS}}U_{\beta,\varepsilon}$, and
the continuum limit gives \cite{Calabrese:2004eu}
\begin{equation}
S_{A}^{\left(0\right)}=\Phi_{\beta,A}=\frac{c}{3}\log\left[\frac{\beta}{\pi\varepsilon}\sinh\frac{\pi L_{A}}{\beta}\right]+o\left(1\right).\label{eq:intro thermal result}
\end{equation}
We verify that the complete thermal worldsheet and spacetime current
reduce to the vacuum construction as $\beta\to\infty$. For $L_{A}/\beta\to\infty$,
the entropy slope approaches 

\begin{equation}
\left.\frac{\partial S_{A}^{(0)}}{\partial L_{A}}\right|_{\beta}\longrightarrow\frac{\pi c}{3\beta}=\frac{\rho_{S}\ell_{\mathrm{AdS}}}{z_{h}},
\end{equation}
which is the BTZ horizon entropy per unit boundary length \cite{Ryu:2006ef,Banados:1992wn}.

These explicit solutions reveal additional information that is not
contained in the value of the RT entropy. In the pure NS--NS background,
the F1 profiles starting on $A$ reach only a bounded part $B_{\mathrm{hit}}\subset\bar{A}$
of the regulated complement, even though their flux computes the entropy
of the reduced state on $A$, which is defined using the full complement.
To understand this feature, we vary the NS--NS fraction $\eta$ in
a classical mixed-flux background \cite{Cagnazzo:2012se,Hoare:2013pma}.
The F1 equation selects the stationary profile 
\begin{equation}
\theta_{\eta}\left(r\right)=\theta_{h}+\frac{\ell_{\mathrm{AdS}}}{r_{h}}\log\frac{\left(1+\eta\right)r}{\mathcal{D}_{\eta}\left(r\right)+\eta r_{h}},\qquad\mathcal{D}_{\eta}^{2}\left(r\right)=\left(1-\eta^{2}\right)r^{2}+\eta^{2}r_{h}^{2}.\label{eq:intro mixed profile}
\end{equation}
At $\eta=0$, the strings follow the geodesics normal to the RT surface,
as in the bit-thread construction \cite{Agon:2018lwq,Mintchev:2022fcp,Caggioli:2024uza}.
At $\eta=1$, Eq. (\ref{eq:intro mixed profile}) becomes the pure
NS--NS profile and all complementary cutoff contacts lie on one side. 

In vacuum, writing $\vartheta=\mathrm{arc}\tanh\eta$ and $U_{\varepsilon}=\mathrm{arc}\cosh\left(R/\varepsilon\right)$,
the fraction of the entropy flux reaching the left complement is 
\begin{equation}
\frac{\Phi_{\mathrm{L}}}{S_{A}^{\left(0\right)}}=\frac{\left[U_{\varepsilon}-\vartheta\right]_{+}}{2U_{\varepsilon}},\qquad\eta_{\mathrm{crit}}=\tanh U_{\varepsilon}=\sqrt{1-\frac{\varepsilon^{2}}{R^{2}}}.\label{eq:intro vacuum routing}
\end{equation}
Thus the F1 dynamics predict the precise finite-cutoff threshold at
which the left flux vanishes.

In planar BTZ, we identify which strings connect the two exteriors
by where they cross the regulated RT surface on the time-zero slice.
We denote the crossing coordinate by $\theta_{h}$, the signed proper
distance from the midpoint of the RT surface in units of $\ell_{\mathrm{AdS}}$.
The string ensemble spans $\theta_{h}\in\left[-U,U\right]$, where
$U=U_{\beta,\varepsilon}$. For $0\leq\eta<1$, the strings that connect
the two exteriors are those with $-d-\vartheta<\theta_{h}<d-\vartheta$,
where $d=L_{A}/\left(2z_{h}\right)$. Their signed flux through the
BTZ bifurcation line is therefore proportional to the length of the
overlap:

\begin{equation}
\Phi_{\mathrm{hor}}^{\mathrm{net}}=\rho_{S}\ell_{\mathrm{AdS}}\mathrm{Length}\left(\left[-U,U\right]\cap\left[-d-\vartheta,d-\vartheta\right]\right).\label{eq:intro net horizon flux}
\end{equation}
When $U\geq d+\vartheta$, this becomes 
\begin{equation}
\Phi_{\mathrm{hor}}^{\mathrm{net}}=2d\rho_{S}\ell_{\mathrm{AdS}}=s_{\mathrm{th}}L_{A}=\frac{\pi cL_{A}}{3\beta}.\label{eq:intro thermal plateau}
\end{equation}
At fixed cutoff the overlap in Eq. (\ref{eq:intro net horizon flux})
eventually shrinks to zero as $\eta$ increases. At the pure NS--NS
point, the net horizon flux vanishes although the interval entropy
and thermal entropy density are unchanged. Some F1 profiles enter
the second exterior and return, and their two bifurcation-line crossings
have opposite orientations and cancel in the signed string charge,
while each profile crosses the interval RT surface only once. This
exhibits the main distinction between the entropy and its physical
F1 realization: the RT surface fixes the total maximal flux, whereas
the worldsheet dynamics determine where that flux is routed.

Finally, we clarify two aspects of this interpretation. First, a positive
reweighting of the string ensemble over the same range of RT crossing
points changes the Kalb--Ramond charge density and the density of
cutoff contacts, but in the pure NS--NS construction it does not
change the normalized entropy flow or its maximal flux. Second, $B_{\mathrm{hit}}$
characterizes the support of one particular F1 maximizer. In other
words, it is not the subsystem traced out in the definition of $\rho_{A}$.
Indeed, in the vacuum we exhibit an interval $C_{0}$ outside $B_{\mathrm{hit}}$
for which 

\begin{equation}
I^{\left(0\right)}\left(A:C_{0}\right)=\frac{c}{3}\log\frac{16}{9}>0.
\end{equation}
Thus the absence of F1 contacts from this maximizer does not imply
the absence of leading boundary correlations.

The paper is organized as follows. In Section \ref{sec:offflow-review},
we review the F1 worldsheet equations, the Kalb--Ramond current,
and its conserved spatial projection. Section \ref{sec:single-interval-F1}
constructs the vacuum worldsheets and derives the vacuum interval
entropy. Section \ref{subsec:thermal} repeats the construction independently
in planar BTZ, derives the thermal interval entropy, and checks the
vacuum and long-interval limits. Section \ref{sec:f1-stationary-routing}
studies the mixed-flux family and derives the vacuum routing threshold
and the net horizon flux. Section \ref{sec:physical-interpretation}
summarizes the results and discusses string reweighting, cutoff contacts,
and their relation to the boundary reduced state. 

\section{Worldsheet equations and the spatial F1 current}

\label{sec:offflow-review}

In this paper, we consider Poincare $\mathrm{AdS}_{3}$ and planar
BTZ supported by pure NS--NS flux, with constant dilaton $\phi=\phi_{0}$
\cite{Banados:1992wn,Maldacena:2000hw,Horowitz:1993jc}. We first
summarize how an on-shell F1 probe defines the spatial current used
below.

We work in the probe limit, keeping $g_{\mathrm{ab}}$, $B_{\mathrm{ab}}$,
and $\phi$ fixed. The F1 sources therefore do not backreact on the
background at the order considered. The action for one F1 probe is
\cite{Polyakov:1981rd,Callan:1985ia,Dabholkar:1990yf}
\begin{equation}
S_{\mathrm{F1}}=-\frac{1}{4\pi\alpha^{\prime}}\int d^{2}\sigma\bigl(\sqrt{-\gamma}\gamma^{mn}\partial_{m}X^{\mathrm{a}}\partial_{n}X^{\mathrm{b}}g_{\mathrm{ab}}+\epsilon^{mn}\partial_{m}X^{\mathrm{a}}\partial_{n}X^{\mathrm{b}}B_{\mathrm{ab}}\bigr),\label{eq:f1 review action}
\end{equation}
where $\gamma_{mn}$ is the worldsheet metric and $\gamma$ its determinant.
We choose $\epsilon^{\tau\sigma}=+1$, with $B$ specified in each
background. In conformal gauge, variation of the F1 action with respect
to $X^{\mathrm{a}}$ gives the embedding equations

\begin{equation}
\mathcal{E}^{\mathrm{a}}-H^{\mathrm{a}}{}_{\mathrm{bc}}\partial_{\tau}X^{\mathrm{b}}\partial_{\sigma}X^{\mathrm{c}}=0,\label{eq:EOM}
\end{equation}
where
\begin{equation}
\mathcal{E}^{\mathrm{a}}\equiv-\partial_{\tau}^{2}X^{\mathrm{a}}+\partial_{\sigma}^{2}X^{\mathrm{a}}+\Gamma^{\mathrm{a}}{}_{\mathrm{bc}}\big(-\partial_{\tau}X^{\mathrm{b}}\partial_{\tau}X^{\mathrm{c}}+\partial_{\sigma}X^{\mathrm{b}}\partial_{\sigma}X^{\mathrm{c}}\big).\label{eq:XI}
\end{equation}
The independent worldsheet-metric equation also imposes
\begin{align}
g_{\mathrm{ab}}\partial_{\tau}X^{\mathrm{a}}\partial_{\sigma}X^{\mathrm{b}} & =0,\nonumber \\
g_{\mathrm{ab}}\big(\partial_{\tau}X^{\mathrm{a}}\partial_{\tau}X^{\mathrm{b}}+\partial_{\sigma}X^{\mathrm{a}}\partial_{\sigma}X^{\mathrm{b}}\big) & =0.\label{eq:constraints}
\end{align}

For $\mathcal{N}$ strings, variation of the total Kalb--Ramond coupling
defines the normalized source current in the convention of Refs. \cite{Wu:2025qwc,Wu:2026pgc}
\begin{align}
\delta_{B}S_{\mathrm{F1}} & =-\frac{1}{4\pi\alpha^{\prime}}\int d^{3}x\sqrt{-g}j^{\mathrm{ab}}\delta B_{\mathrm{ab}},\nonumber \\
j^{\mathrm{ab}}\left(x\right) & =\frac{1}{\sqrt{-g}}\sum_{k=1}^{\mathcal{N}}\int d\tau d\sigma\epsilon^{mn}\partial_{m}X_{k}^{\mathrm{a}}\partial_{n}X_{k}^{\mathrm{b}}\delta^{\left(3\right)}\left(x-X_{k}\left(\tau,\sigma\right)\right).\label{eq:review current}
\end{align}
The current is conserved in the bulk, $\nabla_{\mathrm{a}}j^{\mathrm{ab}}=0$,
for worldsheets without interior endpoints. 

With vanishing shift, let $N_{\mathrm{ADM}}$ be the lapse and $n_{\mu}=\left(-N_{\mathrm{ADM}},0,0\right)$
the future-directed unit normal covector to a constant-time slice.
Its spatial current is 
\begin{equation}
q^{\mu}=-n_{\nu}j^{\nu\mu},\qquad q^{0}=0,\qquad q^{i}=N_{\mathrm{ADM}}j^{0i}.\label{eq:f1 review projection}
\end{equation}
Since $\sqrt{-g}=N_{\mathrm{ADM}}\sqrt{h}$ and $j^{00}=0$, the time
component of bulk current conservation gives 
\begin{equation}
0=\partial_{i}\bigl(\sqrt{-g}j^{i0}\bigr)=-\partial_{i}\bigl(\sqrt{h}q^{i}\bigr).\label{eq:f1 review spatial conservation}
\end{equation}
Thus $\nabla_{i}q^{i}=0$ on each slice, including when $j^{ij}\ne0$.

In both interval calculations, $|q|$ is constant and nonzero inside
the bundle on the slice used to compute the entropy. We then use the
constant rescaling 
\begin{equation}
v_{\mathrm{phys}}^{i}=\rho_{S}\widehat{q}^{i}=\frac{\rho_{S}}{|q|}q^{i},\label{eq:offflow F1}
\end{equation}
on the bundle, with zero flow outside. This preserves conservation
and saturates the norm bound inside the bundle. 

\section{Single-interval entanglement entropy from physical F1 sources}

\label{sec:single-interval-F1}

In this section, we compute the leading entanglement entropy of a
single interval in the CFT$_{2}$ vacuum using on-shell F1 probes
in Poincare AdS$_{3}$. We first construct a family of F1 worldsheets
and choose an ensemble whose intersections with the regulated RT surface
are equally spaced in proper length. We then coarse-grain the discrete
string current, project it onto a constant-time slice, and evaluate
the spatial current at $t=0$. Finally, we normalize this current
using the string-frame entropy capacity and show that the resulting
flow saturates the flux bound on the RT surface, reproducing the standard
vacuum single-interval entropy.

\subsection{Physical F1 strings in Poincare AdS$_{3}$}

\label{subsec:intF1-vacuum-strings}

Now, we consider the boundary interval $A=\left(-R,R\right)$ at $t=0$
in the string-frame background 

\begin{equation}
dS_{\mathrm{AdS}}^{2}=\frac{\ell_{\mathrm{AdS}}^{2}}{z^{2}}\left(-dt^{2}+dx^{2}+dz^{2}\right),\qquad B=-\frac{\ell_{\mathrm{AdS}}^{2}}{z^{2}}dt\wedge dx,\qquad\phi=\phi_{0},
\end{equation}
where $\ell_{\mathrm{AdS}}$ is the AdS radius and $z>0$. In particular, 

\begin{equation}
H=dB=\frac{2\ell_{\mathrm{AdS}}^{2}}{z^{3}}dt\wedge dx\wedge dz,\qquad H_{txz}=\frac{2\ell_{\mathrm{AdS}}^{2}}{z^{3}}.
\end{equation}
To construct a simple family centered on the right endpoint $(0,R,0)$
of the interval, we introduce the dimensionless worldsheet coordinates
\begin{equation}
\tau=\frac{t}{z},\qquad\sigma=\frac{R-x}{z}.
\end{equation}
This ansatz reduces the three embedding functions $t\left(\tau,\sigma\right)$,
$x\left(\tau,\sigma\right)$ and $z\left(\tau,\sigma\right)$ to the
one undetermined function $z\left(\tau,\sigma\right)$:
\begin{equation}
t=\tau z,\qquad x=R-\sigma z.\label{eq:gauge}
\end{equation}
The differentials give
\begin{equation}
dt=zd\tau+\tau dz,\qquad dx=-zd\sigma-\sigma dz.
\end{equation}
Substituting this gauge into $dS_{\mathrm{AdS}}^{2}$, its induced
line element is
\begin{equation}
ds_{\mathrm{ws}}^{2}=\ell_{\mathrm{AdS}}^{2}\left(-d\tau^{2}+d\sigma^{2}+\frac{d\mathcal{D}}{z}dz+\frac{\mathcal{D}}{z^{2}}dz^{2}\right),
\end{equation}
where $\mathcal{D}\equiv1+\sigma^{2}-\tau^{2}$. The last two terms
can be rewritten as
\begin{equation}
\frac{d\mathcal{D}}{z}dz+\frac{\mathcal{D}}{z^{2}}dz^{2}=\frac{dz}{z^{2}}d\left(\mathcal{D}z\right).
\end{equation}
We select the nonconstant branch with a flat induced metric,
\begin{equation}
ds_{\mathrm{ws}}^{2}=\ell_{\mathrm{AdS}}^{2}\left(-d\tau^{2}+d\sigma^{2}\right),
\end{equation}
This is a convenient conformal-gauge solution of the worldsheet metric
constraints. The embedding equations must still be checked independently.
On the branch $dz\neq0$, the required cancellation is equivalent
to
\begin{equation}
d\left(\mathcal{D}z\right)=0
\end{equation}
Writing the positive integration constant as $2z_{0}$, so that $\mathcal{D}z=2z_{0}$,
and restricting to $z_{0}>0$ and $\mathcal{D}>0$, we obtain
\begin{equation}
X^{t}\left(\tau,\sigma\right)=\frac{2z_{0}\tau}{\mathcal{D}},\qquad X^{x}\left(\tau,\sigma\right)=R-\frac{2z_{0}\sigma}{\mathcal{D}},\qquad X^{z}\left(\tau,\sigma\right)=\frac{2z_{0}}{\mathcal{D}}.\label{eq:embed solution}
\end{equation}
The conditions $z_{0}>0$ and $\mathcal{D}>0$ place the solution
in the Poincare patch $z>0$. This verifies the constraints in Eq.
(\ref{eq:constraints}). Choosing the Weyl representative in which
the independent metric equals the induced metric gives
\begin{align}
\gamma_{mn} & =g_{\mathrm{ab}}\partial_{m}X^{\mathrm{a}}\partial_{n}X^{\mathrm{b}}=\ell_{\mathrm{AdS}}^{2}\mathrm{diag}\left(-1,1\right),\nonumber \\
\sqrt{-\det\gamma} & =\ell_{\mathrm{AdS}}^{2},\qquad\sqrt{-\gamma}\gamma^{mn}=\mathrm{diag}\left(-1,1\right).
\end{align}

To identify the target space surface swept out by this embedding,
use Eq. (\ref{eq:gauge}):

\begin{equation}
t^{2}=\tau^{2}z^{2},\qquad\left(x-R\right)^{2}=\sigma^{2}z^{2}.
\end{equation}
Combining them and adding $z^{2}$ to both sides gives

\begin{equation}
\left(x-R\right)^{2}+z^{2}-t^{2}=z^{2}\left(1+\sigma^{2}-\tau^{2}\right).\label{eq:surface}
\end{equation}
Using (\ref{eq:embed solution}), we have
\begin{equation}
\left(x-R\right)^{2}+z^{2}-t^{2}=2z_{0}z.\label{eq:circle}
\end{equation}
Equivalently,

\begin{equation}
\left(x-R\right)^{2}+\left(z-z_{0}\right)^{2}-t^{2}=z_{0}^{2},\label{eq:worldsheet in vacuum}
\end{equation}
This equation is the target space image of the string worldsheet described
by $t\left(\tau,\sigma\right)$, $x\left(\tau,\sigma\right)$ and
$z\left(\tau,\sigma\right)$.

Now, we can check whether this configuration (\ref{eq:embed solution})
satisfies the complete F1 equations (\ref{eq:EOM}) and reproduces
(\ref{eq:worldsheet in vacuum}). Substituting the solution (\ref{eq:embed solution}),
\begin{equation}
X^{t}\left(\tau,\sigma\right)\equiv t\left(\tau,\sigma\right)=\frac{2z_{0}\tau}{\mathcal{D}},\qquad X^{x}\left(\tau,\sigma\right)\equiv x\left(\tau,\sigma\right)=R-\frac{2z_{0}\sigma}{\mathcal{D}},\qquad X^{z}\left(\tau,\sigma\right)\equiv z\left(\tau,\sigma\right)=\frac{2z_{0}}{\mathcal{D}},\label{eq:definition gauge}
\end{equation}
into the equations of motion $\mathcal{E}^{\mathrm{a}}-H^{\mathrm{a}}{}_{\mathrm{bc}}\partial_{\tau}X^{\mathrm{b}}\partial_{\sigma}X^{\mathrm{c}}$,
we find

\begin{equation}
\mathcal{E}^{\mathrm{a}}-H^{\mathrm{a}}{}_{\mathrm{bc}}\partial_{\tau}X^{\mathrm{b}}\partial_{\sigma}X^{\mathrm{c}}=-\frac{2z}{z_{0}}\left(t,x-R,z-z_{0}\right)^{\mathrm{a}}+\frac{2z}{z_{0}}\left(t,x-R,z-z_{0}\right)^{\mathrm{a}}=0.
\end{equation}
The embedding therefore satisfies the complete F1 equations and the
worldsheet metric constraints. Moreover, at $t=0$, we have 
\begin{equation}
\left(x-R\right)^{2}+\left(z-z_{0}\right)^{2}=z_{0}^{2},
\end{equation}
which is a circle tangent to the ideal boundary at $\left(R,0\right)$.
The same family of curves appears in the geometric construction of
maximal flows in Ref. \cite{Hubeny:2018bri}. In this paper, it arises
as the time-zero section of on-shell F1 worldsheets. The RT surface
is \cite{Ryu:2006bv,Ryu:2006ef}
\begin{equation}
\gamma_{A}:\qquad x^{2}+z^{2}=R^{2}.\label{eq:RT}
\end{equation}
At their intersection in the bulk, Eq. (\ref{eq:surface}) implies
$z_{0}z=R\left(R-x\right)$. Their Euclidean normal vectors obey 
\begin{equation}
\left(x,z\right)\cdot\left(x-R,z-z_{0}\right)=R^{2}-Rx-z_{0}z=0.
\end{equation}
Since the spatial string-frame metric is conformal to $dx^{2}+dz^{2}$,
the curves also intersect orthogonally in the physical metric.

\subsection{The physical string current and vacuum interval entropy}

\label{subsec:intF1-vacuum-entropy}

To define the spacing between adjacent strings, we use a geometric
regulator that keeps the ideal endpoints $\pm R$ fixed and truncates
their surface at $z=\varepsilon$:

\begin{equation}
a_{\varepsilon}=\sqrt{R^{2}-\varepsilon^{2}}<R,\qquad A_{\varepsilon}=\left\{ z=\varepsilon,\ -a_{\varepsilon}\leq x\leq a_{\varepsilon}\right\} ,\qquad0<\varepsilon<R.\label{eq:regulator}
\end{equation}
At $t=0$, each string attached to $A_{\varepsilon}$ satisfies (\ref{eq:circle}):

\begin{equation}
\left(x-R\right)^{2}+\varepsilon^{2}=2z_{0}\varepsilon.
\end{equation}
Thus, the string passing through the cutoff point $x$ has integration
constant

\begin{equation}
z_{0}\left(x\right)=\frac{\left(x-R\right)^{2}+\varepsilon^{2}}{2\varepsilon}.
\end{equation}
Using $a_{\varepsilon}^{2}+\varepsilon^{2}=R^{2}$, we have
\begin{eqnarray}
z_{0,\min} & = & \frac{\left(R-a_{\varepsilon}\right)^{2}+\varepsilon^{2}}{2\varepsilon}=\frac{R\left(R-a_{\varepsilon}\right)}{\varepsilon},\nonumber \\
z_{0,\max} & = & \frac{\left(R+a_{\varepsilon}\right)^{2}+\varepsilon^{2}}{2\varepsilon}=\frac{R\left(R+a_{\varepsilon}\right)}{\varepsilon}.
\end{eqnarray}
The endpoint values determine the logarithmic size of the range:

\begin{equation}
\frac{z_{0,\max}}{z_{0,\min}}=\frac{R+a_{\varepsilon}}{R-a_{\varepsilon}}=\frac{\left(R+a_{\varepsilon}\right)^{2}}{\varepsilon^{2}},
\end{equation}
due to $\left(R-a_{\varepsilon}\right)\left(R+a_{\varepsilon}\right)=R^{2}-a_{\varepsilon}^{2}=\varepsilon^{2}$.
Therefore, we can define the notation

\begin{equation}
U_{\varepsilon}\equiv\frac{1}{2}\log\frac{z_{0,\max}}{z_{0,\min}}=\log\frac{R+\sqrt{R^{2}-\varepsilon^{2}}}{\varepsilon}=\mathrm{arc}\cosh\frac{R}{\varepsilon}.\label{eq:U}
\end{equation}
Since $z_{0,\min}z_{0,\max}=R^{2}$, the endpoints can equivalently
be expressed as

\begin{equation}
z_{0,\min}=Re^{-U_{\varepsilon}},\qquad z_{0,\max}=Re^{U_{\varepsilon}}.
\end{equation}
Since $R$ is the geometric mean of $z_{0,\min}$ and $z_{0,\max}$,
we use it as the reference scale for the dimensionless string label
\begin{equation}
u=\log\frac{z_{0}}{R},\qquad-U_{\varepsilon}\leq u\leq U_{\varepsilon}.\label{eq:F1 label}
\end{equation}
 Thus $U_{\varepsilon}$ is the half-width of the allowed range of
$u$. The F1 equations determine the string profile for each $u$.
To form the ensemble, we choose equally spaced values of $u$. This
places the intersections of the strings with the RT surface at equal
intervals of proper length.

\begin{figure}[h]
\begin{centering}
\includegraphics[scale=1.4]{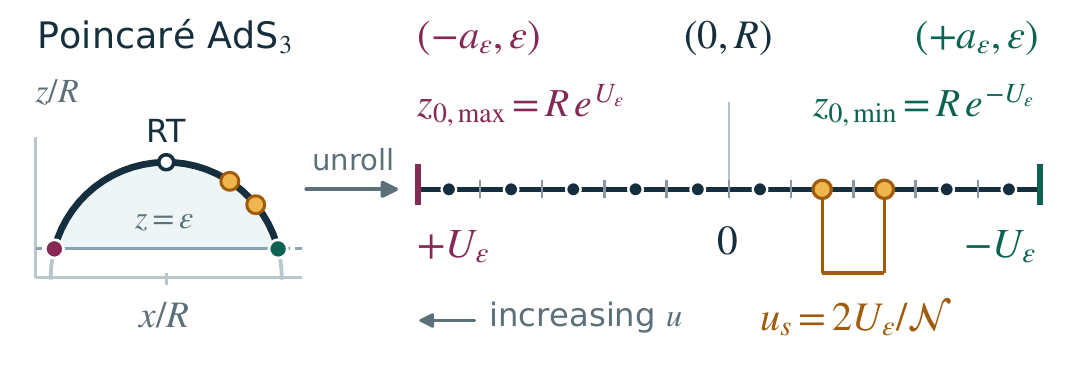}
\par\end{centering}
\caption{\label{fig:vaccum ruler}The regulated vacuum RT surface in Poincare
AdS$_{3}$ at $t=0$ is shown on the left and unfolded onto the arclength
ruler on the right. The left and right cutoff endpoints map to $+U_{\varepsilon}$
and $-U_{\varepsilon}$, corresponding to $z_{0,\max}$ and $z_{0,\min}$,
respectively. The dots on the ruler mark F1 crossings at the cell
midpoints. The same two adjacent crossings are highlighted in gold
in both panels. Their label separation $u_{s}$ corresponds to a proper
distance $\ell_{\mathrm{AdS}}u_{s}$ along the surface.}
\end{figure}

The RT surface and an F1 profile satisfy

\begin{equation}
\mathrm{RT}:\quad x^{2}+z^{2}=R^{2},\qquad\mathrm{F1}:\quad\left(x-R\right)^{2}+z^{2}=2z_{0}z.
\end{equation}
Solving Eqs. (\ref{eq:circle}) and (\ref{eq:RT}) at $t=0$ gives
the bulk intersection of the F1 string profile with the RT surface:
\begin{equation}
x=R\frac{R^{2}-z_{0}^{2}}{R^{2}+z_{0}^{2}},\qquad z=\frac{2R^{2}z_{0}}{R^{2}+z_{0}^{2}}.
\end{equation}
Substituting $z_{0}=Re^{u}$, we obtain
\begin{equation}
x=-R\tanh u,\qquad z=R\mathrm{sech}u.
\end{equation}
Since $dx^{2}+dz^{2}=R^{2}\mathrm{sech}^{2}udu^{2}=z^{2}du^{2}$,
the proper line element along the RT curve is 

\begin{equation}
dA_{S}=\frac{\ell_{\mathrm{AdS}}}{z}\sqrt{dx^{2}+dz^{2}}=\frac{\ell_{\mathrm{AdS}}}{z}\sqrt{R^{2}\mathrm{sech}^{2}udu^{2}}=\ell_{\mathrm{AdS}}\left|du\right|.
\end{equation}
Thus, the equal spacing in $u$ gives equal proper spacing along the
RT curve. This parallels the BTZ construction in our previous work
\cite{Wu:2026pgc}, where equal angular spacing gives equal proper
spacing along the horizon. Based on this observation, we can follow
the notation of the angular separation $\theta_{s}=2\pi/\mathcal{N}$
in the BTZ case and define the separation between adjacent strings
by
\begin{equation}
u_{s}=\frac{2U_{\varepsilon}}{\mathcal{N}}.
\end{equation}
Divide $\left[-U_{\varepsilon},U_{\varepsilon}\right]$ into $\mathcal{N}$
equal cells and place one string at the midpoint of each cell:

\begin{equation}
u_{k}=-U_{\varepsilon}+\left(k+\frac{1}{2}\right)u_{s},\qquad k=0,\ldots,\mathcal{N}-1.
\end{equation}
The corresponding integration constants are 
\begin{equation}
z_{0,k}=Re^{u_{k}}=z_{0,\min}\exp\left[\left(k+\frac{1}{2}\right)u_{s}\right],
\end{equation}
and the adjacent strings satisfy
\begin{equation}
\log\frac{z_{0,k+1}}{z_{0,k}}=u_{s}.
\end{equation}
The regulator selects this ensemble through its intersections with
$A_{\varepsilon}$ at $t=0$. Each $u_{k}$ is then a fixed label
of a complete on-shell worldsheet at all times. Fig. (\ref{fig:vaccum ruler})
shows the limiting values of $z_{0}$ at the two cutoff endpoints
and how midpoint sampling gives equal proper spacing between F1 crossings
along the regulated RT surface. 

We now return to the normalized current of $\mathcal{N}$ equally
oriented strings: 
\begin{equation}
j^{\mathrm{ab}}\left(x\right)=\frac{1}{\sqrt{-g}}\sum_{k=0}^{\mathcal{N}-1}\int d\tau d\sigma\left(\partial_{\tau}X_{k}^{\mathrm{a}}\partial_{\sigma}X_{k}^{\mathrm{b}}-\partial_{\sigma}X_{k}^{\mathrm{a}}\partial_{\tau}X_{k}^{\mathrm{b}}\right)\delta^{\left(3\right)}\left(x-X_{k}\right).\label{eq:vacuum source}
\end{equation}
This source current comes entirely from the Kalb--Ramond coupling
and is independent of the chosen Weyl representative of the auxiliary
worldsheet metric. To simplify this current, we recall the embedding
(\ref{eq:embed solution}), before restricting to $t=0$:
\begin{equation}
X_{k}^{t}=\frac{2z_{0,k}\tau}{\mathcal{D}},\qquad X_{k}^{x}=R-\frac{2z_{0,k}\sigma}{\mathcal{D}},\qquad X_{k}^{z}=\frac{2z_{0,k}}{\mathcal{D}},\qquad\mathcal{D}=1+\sigma^{2}-\tau^{2}>0.
\end{equation}
The inverse map of the embedding family is

\begin{equation}
\tau_{*}=\frac{t}{z},\qquad\sigma_{*}=\frac{R-x}{z},\qquad u=\log\frac{\Delta_{A}}{2Rz},\label{eq:inverse}
\end{equation}
where $\Delta_{A}\equiv(x-R)^{2}+z^{2}-t^{2}$ is a spacetime function.
On the $k$th worldsheet, $\Delta_{A}=2z_{0,k}z$, so the inverse
map returns $u=u_{k}=\log\left(z_{0,k}/R\right)$. The quantities
$\tau_{*}$ and $\sigma_{*}$ are the worldsheet coordinates of the
event $\left(t,x,z\right)$ on that string. The target space delta
distribution of (\ref{eq:vacuum source}) fixes $\tau=\tau_{*}$ and
$\sigma=\sigma_{*}$. Allowing $u$ to vary between different members
of the family, and using $z_{0}=Re^{u}$ and $z=2z_{0}/\mathcal{D}$,
we have
\begin{equation}
\frac{dz}{z}=du-\frac{d\mathcal{D}}{\mathcal{D}}.
\end{equation}
Combining with Eq. (\ref{eq:gauge}) gives
\begin{equation}
dt=zd\tau+\tau dz,\qquad dx=-zd\sigma-\sigma dz.
\end{equation}
or equivalently,
\begin{equation}
dt\wedge dx\wedge dz=-z^{2}d\tau\wedge d\sigma\wedge dz=-z^{3}d\tau\wedge d\sigma\wedge du.
\end{equation}
Therefore the absolute Jacobian is $z^{3}$, and the target space
delta distribution becomes 
\begin{equation}
\delta^{\left(3\right)}\left(x-X_{k}\right)=\frac{1}{z^{3}}\delta\left(\tau-\tau_{*}\right)\delta\left(\sigma-\sigma_{*}\right)\delta\left(u-u_{k}\right).
\end{equation}
Thus, the $\mathcal{N}$ equally oriented strings can be simplified
as
\begin{equation}
j^{\mathrm{ab}}=\frac{1}{\sqrt{-g}z^{3}}\left(\partial_{\tau}X^{\mathrm{a}}\partial_{\sigma}X^{\mathrm{b}}-\partial_{\sigma}X^{\mathrm{a}}\partial_{\tau}X^{\mathrm{b}}\right)_{*}\sum_{k=0}^{\mathcal{N}-1}\delta\left(u-u_{k}\right),
\end{equation}
where subscript $*$ denotes that the worldsheet derivatives are taken
at fixed $u$ and then evaluated at $\left(\tau_{*},\sigma_{*},u\left(x\right)\right)$.
Substituting the embedding solution (\ref{eq:embed solution}),
\begin{align}
\partial_{\tau}X^{t}\partial_{\sigma}X^{x}-\partial_{\sigma}X^{t}\partial_{\tau}X^{x} & =\frac{z^{2}\left(\mathcal{D}-2\right)}{\mathcal{D}},\nonumber \\
\partial_{\tau}X^{t}\partial_{\sigma}X^{z}-\partial_{\sigma}X^{t}\partial_{\tau}X^{z} & =-\frac{2\sigma z^{2}}{\mathcal{D}},\nonumber \\
\partial_{\tau}X^{x}\partial_{\sigma}X^{z}-\partial_{\sigma}X^{x}\partial_{\tau}X^{z} & =\frac{2\tau z^{2}}{\mathcal{D}}.
\end{align}
Using Eq. (\ref{eq:inverse}), $\mathcal{D}=\Delta_{A}/z^{2}$ and
$\sqrt{-g}=\ell_{\mathrm{AdS}}^{3}/z^{3}$, the nonvanishing discrete
current becomes 
\begin{align}
j^{0x} & =\frac{z^{2}\left[\left(x-R\right)^{2}-z^{2}-t^{2}\right]}{\ell_{\mathrm{AdS}}^{3}\Delta_{A}}\sum_{k=0}^{\mathcal{N}-1}\delta\left(u-u_{k}\right),\nonumber \\
j^{0z} & =\frac{2z^{3}\left(x-R\right)}{\ell_{\mathrm{AdS}}^{3}\Delta_{A}}\sum_{k=0}^{\mathcal{N}-1}\delta\left(u-u_{k}\right),\nonumber \\
j^{xz} & =\frac{2tz^{3}}{\ell_{\mathrm{AdS}}^{3}\Delta_{A}}\sum_{k=0}^{\mathcal{N}-1}\delta\left(u-u_{k}\right).
\end{align}

Next, we need to coarse-grain the discrete sum. Since $u$ is nonperiodic,
we use ordinary delta functions and write 
\begin{align}
\sum_{k=0}^{\mathcal{N}-1}\delta\left(u-u_{k}\right) & =\frac{1}{u_{s}}\left[u_{s}\sum_{k=0}^{\mathcal{N}-1}\delta\left(u-u_{k}\right)\right]\nonumber \\
 & =\frac{\mathcal{N}}{2U_{\varepsilon}}\left[u_{s}\sum_{k=0}^{\mathcal{N}-1}\delta\left(u+U_{\varepsilon}-\left(k+\frac{1}{2}\right)u_{s}\right)\right].
\end{align}
For a smooth test function $f$, the term in brackets satisfies 
\begin{equation}
u_{s}\sum_{k=0}^{\mathcal{N}-1}f\left(u_{k}\right)\longrightarrow\int_{-U_{\varepsilon}}^{U_{\varepsilon}}f\left(u\right)du,
\end{equation}
by the midpoint Riemann-sum limit. Its coarse-grained value is therefore
one inside the range of string labels. Consequently, 
\begin{equation}
\sum_{k=0}^{\mathcal{N}-1}\delta\left(u-u_{k}\right)\xrightarrow{\mathrm{coarse\text{-}graining}}\frac{\mathcal{N}}{2U_{\varepsilon}},\qquad-U_{\varepsilon}<u<U_{\varepsilon}.
\end{equation}
Substituting this coarse-grained density into the discrete current
gives the smooth current at general target-space time:
\begin{equation}
j^{0x}=\frac{\mathcal{N}z^{2}\left[\left(x-R\right)^{2}-z^{2}-t^{2}\right]}{2U_{\varepsilon}\ell_{\mathrm{AdS}}^{3}\Delta_{A}},\qquad j^{0z}=\frac{\mathcal{N}z^{3}\left(x-R\right)}{U_{\varepsilon}\ell_{\mathrm{AdS}}^{3}\Delta_{A}},\qquad j^{xz}=\frac{\mathcal{N}tz^{3}}{U_{\varepsilon}\ell_{\mathrm{AdS}}^{3}\Delta_{A}},
\end{equation}
which hold for $\Delta_{A}>0$ and $-U_{\varepsilon}<u<U_{\varepsilon}$. 

\begin{figure}[h]
\begin{centering}
\includegraphics[scale=0.18]{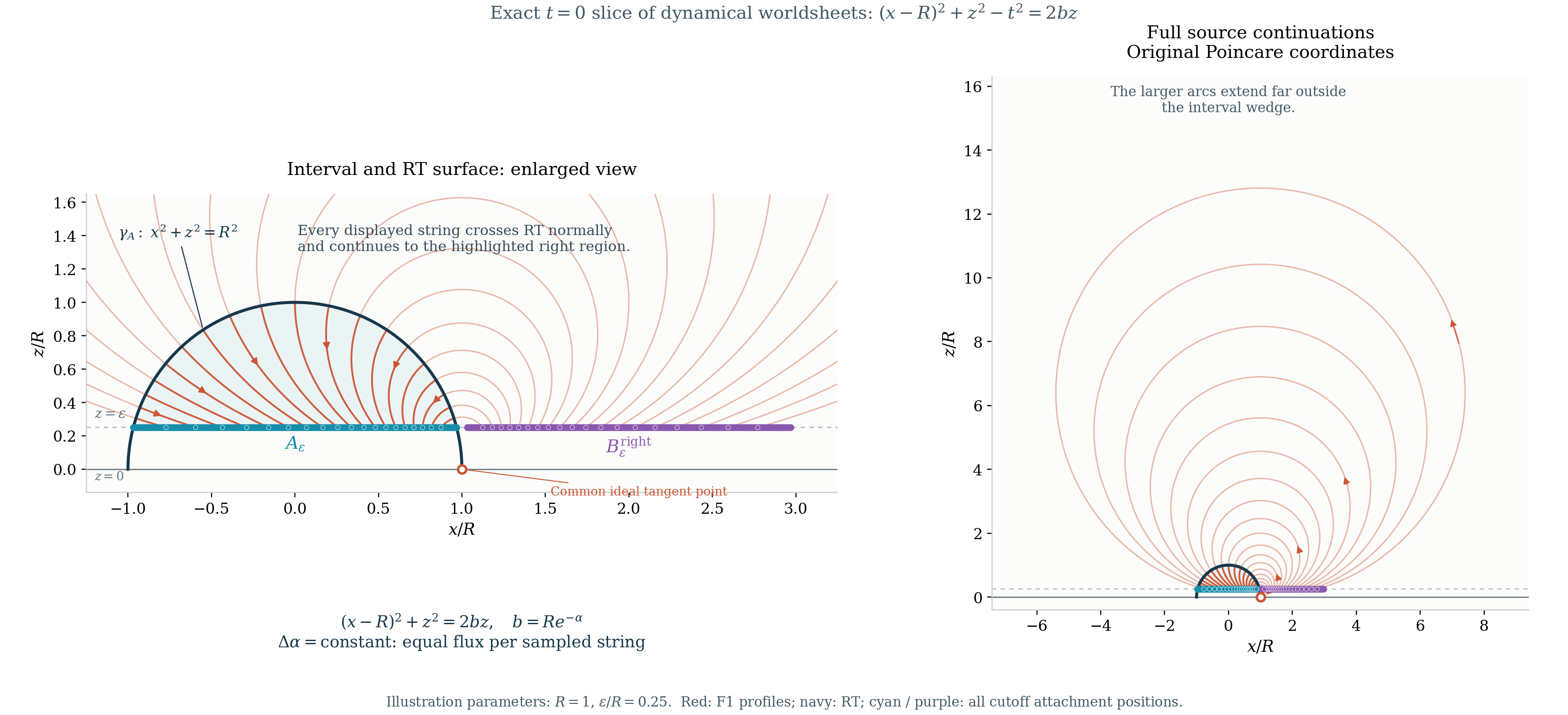}
\par\end{centering}
\caption{\label{fig:vaccum}The red curves are the $t=0$ profiles $\left(x-R\right)^{2}+z^{2}=2z_{0}z$,
sampled at equal intervals of $u=\log\left(z_{0}/R\right)$. They
cross the RT surface $\gamma_{A}$ (navy) orthogonally, with successive
intersections equally spaced in proper length. The left panel enlarges
the region around the interval and its RT surface. The right panel
shows how the same profiles continue outside that region. Cyan marks
the regulated interval $A_{\varepsilon}$, and violet marks the part
of its complement reached by the string bundle at $z=\varepsilon$.
Arrows indicate the F1 charge orientation from the complement toward
$A_{\varepsilon}$. The plot uses $R=1$ and $\varepsilon/R=0.25$.}
\end{figure}

We now project onto a constant-$t$ slice. The lapse and spatial metric
are 
\begin{equation}
N_{\mathrm{ADM}}=\frac{\ell_{\mathrm{AdS}}}{z},\qquad h_{ij}dx^{i}dx^{j}=\frac{\ell_{\mathrm{AdS}}^{2}}{z^{2}}(dx^{2}+dz^{2}),\qquad\sqrt{h}=\frac{\ell_{\mathrm{AdS}}^{2}}{z^{2}}.
\end{equation}
The future-directed unit normal has covariant components $n_{\mu}=\left(-\ell_{\mathrm{AdS}}/z,0,0\right)$.
Thus 
\begin{equation}
q^{\mu}=-n_{\nu}j^{\nu\mu},\qquad q^{0}=0,\qquad q^{i}=N_{\mathrm{ADM}}j^{0i},\quad i=x,z,
\end{equation}
which yields 
\begin{equation}
q^{x}=\frac{\mathcal{N}z\left[\left(x-R\right)^{2}-z^{2}-t^{2}\right]}{2U_{\varepsilon}\ell_{\mathrm{AdS}}^{2}\Delta_{A}},\qquad q^{z}=\frac{\mathcal{N}z^{2}\left(x-R\right)}{U_{\varepsilon}\ell_{\mathrm{AdS}}^{2}\Delta_{A}}.
\end{equation}
The current is conserved on every constant time slice:

\begin{equation}
\partial_{x}\left(\sqrt{h}q^{x}\right)=-\partial_{z}\left(\sqrt{h}q^{z}\right)=\frac{2\mathcal{N}z\left(x-R\right)}{U_{\varepsilon}\Delta_{A}^{2}},\qquad\nabla_{i}q^{i}=0.
\end{equation}
Its norm at general time is 
\begin{align}
|q|^{2} & =\frac{\ell_{\mathrm{AdS}}^{2}}{z^{2}}\left[\left(q^{x}\right)^{2}+\left(q^{z}\right)^{2}\right]\nonumber \\
 & =\left(\frac{\mathcal{N}}{2\ell_{\mathrm{AdS}}U_{\varepsilon}}\right)^{2}\frac{\left[\left(x-R\right)^{2}-z^{2}-t^{2}\right]^{2}+4z^{2}\left(x-R\right)^{2}}{\Delta_{A}^{2}}\nonumber \\
 & =\left(\frac{\mathcal{N}}{2\ell_{\mathrm{AdS}}U_{\varepsilon}}\right)^{2}\left(1+\frac{4t^{2}z^{2}}{\Delta_{A}^{2}}\right),\label{eq:general norm}
\end{align}
where the final equality uses 
\begin{equation}
\left[\left(x-R\right)^{2}-z^{2}-t^{2}\right]^{2}+4z^{2}\left(x-R\right)^{2}=\Delta_{A}^{2}+4t^{2}z^{2}.
\end{equation}
Note that the ADM projection gives a spatial current on each constant-$t$
slice, with $t$ labeling the slice. Since the boundary interval $A=\left(-R,R\right)$
is defined at $t=0$ and its RT surface lies on the same slice, we
now set $t=0$ to compute the entropy flux. On this slice, the projected
current has constant norm and crosses the RT surface orthogonally.
We utilize the notation
\begin{equation}
D_{A}\equiv\left.\Delta_{A}\right|_{t=0}=\left(x-R\right)^{2}+z^{2}.
\end{equation}
The nonvanishing spatial charge density and its norm reduce to 
\begin{equation}
q^{x}=\frac{\mathcal{N}z\left[\left(x-R\right)^{2}-z^{2}\right]}{2U_{\varepsilon}\ell_{\mathrm{AdS}}^{2}D_{A}},\qquad q^{z}=\frac{\mathcal{N}z^{2}\left(x-R\right)}{U_{\varepsilon}\ell_{\mathrm{AdS}}^{2}D_{A}},\qquad|q|=\frac{\mathcal{N}}{2\ell_{\mathrm{AdS}}U_{\varepsilon}}.
\end{equation}
Fig. (\ref{fig:vaccum}) shows representative profiles of the regulated
F1 bundle on the $t=0$ slice and their crossings of the RT surface
$\gamma_{A}$. 
 Thus, the outward unit normal and proper line element on the cutoff
are 
\begin{equation}
n_{\mathrm{cut}}=-\frac{\varepsilon}{\ell_{\mathrm{AdS}}}\partial_{z},\qquad dA_{S}=\frac{\ell_{\mathrm{AdS}}}{\varepsilon}dx.
\end{equation}
The charge flux through the full regulated interval is therefore 
\begin{align}
\Phi_{A}\left[q\right] & \equiv\int_{A_{\varepsilon}}q\cdot n_{\mathrm{cut}}dA_{S}=\frac{\mathcal{N}}{U_{\varepsilon}}\int_{-a_{\varepsilon}}^{a_{\varepsilon}}\frac{R-x}{\left(R-x\right)^{2}+\varepsilon^{2}}dx\nonumber \\
 & =\frac{\mathcal{N}}{2U_{\varepsilon}}\log\frac{\left(R+a_{\varepsilon}\right)^{2}+\varepsilon^{2}}{\left(R-a_{\varepsilon}\right)^{2}+\varepsilon^{2}}=\mathcal{N},\label{eq:charge flux}
\end{align}
where the last step follows from Eq. (\ref{eq:U}).

Following our previous method \cite{Wu:2026pgc}, we next define the
unit-normalized spatial current inside the bundle by 
\begin{equation}
\widehat{q}^{i}=\frac{q^{i}}{|q|},\qquad\widehat{q}=\frac{z}{\ell_{\mathrm{AdS}}D_{A}}\left(\left[\left(x-R\right)^{2}-z^{2}\right]\partial_{x}+2z\left(x-R\right)\partial_{z}\right).
\end{equation}
Since $|q|$ is constant on this slice,
\begin{equation}
|\widehat{q}|=1,\qquad\nabla_{i}\widehat{q}^{i}=\frac{\nabla_{i}q^{i}}{|q|}=0.
\end{equation}
Using Eqs. (\ref{eq:charge flux}) and (\ref{eq:U}), its flux is

\begin{equation}
\Phi_{A}\left[\widehat{q}\right]=\frac{\Phi_{A}\left[q\right]}{|q|}=2\ell_{\mathrm{AdS}}U_{\varepsilon}=2\ell_{\mathrm{AdS}}\mathrm{arc}\cosh\frac{R}{\varepsilon}.
\end{equation}
Thus the normalized geometric flux is independent of $\mathcal{N}$.
We now recall the string-frame norm bound obtained from the off-shell
sphere and replica construction \cite{Wu:2026pgc}, 
\begin{equation}
\rho_{S}=\frac{e^{-2\phi_{0}}}{4G_{N}^{\left(3\right)}}.\label{eq:F1 capacity}
\end{equation}
Since the dilaton is constant, $\rho_{S}$ is constant. Combining
this capacity with the unit current gives the physical entropy flow
\begin{align}
v_{\mathrm{phys}}^{i} & \equiv\rho_{S}\widehat{q}^{i}=\rho_{S}\frac{q^{i}}{|q|},\nonumber \\
v_{\mathrm{phys}} & =\frac{\rho_{S}z}{\ell_{\mathrm{AdS}}D_{A}}\left(\left[\left(x-R\right)^{2}-z^{2}\right]\partial_{x}+2z\left(x-R\right)\partial_{z}\right).\label{eq:physical flow}
\end{align}
Consequently, 
\begin{equation}
\nabla_{i}v_{\mathrm{phys}}^{i}=\rho_{S}\nabla_{i}\widehat{q}^{i}=0,\qquad|v_{\mathrm{phys}}|=\rho_{S}|\widehat{q}|=\rho_{S},\label{eq:flow properties}
\end{equation}
throughout the string bundle. Its entropy flux is 
\begin{equation}
\Phi_{A}\equiv\Phi_{A}\left[v_{\mathrm{phys}}\right]=\rho_{S}\Phi_{A}\left[\widehat{q}\right]=2\rho_{S}\ell_{\mathrm{AdS}}\mathrm{arc}\cosh\frac{R}{\varepsilon}.\label{eq:vacuum flux}
\end{equation}
With central charge $c=6\rho_{S}\ell_{\mathrm{AdS}}=3\ell_{\mathrm{AdS}}e^{-2\phi_{0}}/\left(2G_{N}^{\left(3\right)}\right)$,
this becomes 

\begin{equation}
S_{A}^{\left(0\right)}=\Phi_{A}=\frac{c}{3}\mathrm{arc}\cosh\frac{R}{\varepsilon}=\frac{c}{3}\log\frac{2R}{\varepsilon}+O\left(\frac{\varepsilon^{2}}{R^{2}}\right).
\end{equation}
The finite-cutoff terms refer to the geometric regulator in Eq. (\ref{eq:regulator}).
In the continuum limit, this agrees with the vacuum interval entanglement
entropy of CFT$_{2}$ \cite{Holzhey:1994we,Calabrese:2004eu}, now
obtained from physical F1 sources. This is the vacuum entanglement
entropy between $A$ and its full complement $\bar{A}$. At finite
cutoff, the F1 profiles connect $A_{\varepsilon}$ only to a subinterval
$B_{\mathrm{hit},\varepsilon}$ of the regulated complement. In Section
\ref{sec:f1-stationary-routing}, we show how the F1 equations determine
which parts of the complement are reached as the NS-NS flux fraction
varies, even though the total RT flux remains fixed.

Finally, it remains to verify maximality. The RT surface and its unit
normal pointing into the interval homology region are
\begin{equation}
\gamma_{A}:\ x^{2}+z^{2}=R^{2},\qquad n_{A}=-\frac{z}{\ell_{\mathrm{AdS}}R}\left(x\partial_{x}+z\partial_{z}\right).\label{eq:vacuum RT normal}
\end{equation}
Substitution into Eq. (\ref{eq:physical flow}) gives 
\begin{equation}
v_{\mathrm{phys}}|_{\gamma_{A}}=\rho_{S}n_{A}.
\end{equation}
The flow is therefore normal to the bottleneck and saturates its capacity.
The source intersections with this surface also satisfy 
\begin{equation}
x=R\frac{R^{2}-z_{0}^{2}}{R^{2}+z_{0}^{2}},\qquad z=\frac{2R^{2}z_{0}}{R^{2}+z_{0}^{2}},\qquad dA_{S}=\ell_{\mathrm{AdS}}\frac{|dz_{0}|}{z_{0}}.
\end{equation}
Thus $\mathrm{Length}\left(\gamma_{A,\varepsilon}\right)=2\ell_{\mathrm{AdS}}U_{\varepsilon}$,
and the ensemble is uniform per proper RT length. For any admissible
flow $w^{i}$ in the same regulated homology region, 
\begin{equation}
\nabla_{i}w^{i}=0,\qquad|w|\leq\rho_{S}.
\end{equation}
Flux conservation and the weighted flow--cut bound \cite{Headrick:2017ucz,Wu:2026pgc}
then give 
\begin{align}
\Phi_{A}\left[w\right] & =\int_{\gamma_{A,\varepsilon}}w\cdot n_{A}dA_{S}\leq\int_{\gamma_{A,\varepsilon}}|w|dA_{S}\nonumber \\
 & \leq\rho_{S}\mathrm{Length}\left(\gamma_{A,\varepsilon}\right)=2\rho_{S}\ell_{\mathrm{AdS}}U_{\varepsilon}=\Phi_{A}\left[v_{\mathrm{phys}}\right].
\end{align}
This proves that the entropy flux calculated from the physical string
current is maximal.

\section{Thermal entanglement entropy from physical F1 sources}

\label{subsec:thermal}

In this section, we extend the construction to a single interval in
a thermal state of the CFT$_{2}$ on the infinite line, described
holographically by planar BTZ. We construct on-shell F1 worldsheets
in this background and choose their labels so that the crossings are
equally spaced in proper length along the regulated thermal RT surface.
After coarse-graining the string current and projecting it onto the
$t_{B}=0$ slice, we apply the same string-frame normalization as
in the vacuum case. The flow again saturates the RT flux bound, yielding
the standard single-interval entanglement entropy at inverse temperature
$\beta$.

\subsection{Physical F1 strings in planar BTZ}

\label{subsec:thermal-worldsheets}

We now consider the boundary interval $A=\left(x_{1},x_{2}\right)$
at $t_{B}=0$ in the thermal state on an infinite line. The string-frame
planar BTZ background is \cite{Banados:1992wn,Horowitz:1993jc}
\begin{align}
dS_{\mathrm{BTZ}}^{2} & =\frac{\ell_{\mathrm{AdS}}^{2}}{z_{B}^{2}}\left(-f_{B}dt_{B}^{2}+dx_{B}^{2}+\frac{dz_{B}^{2}}{f_{B}}\right),\qquad f_{B}=1-\frac{z_{B}^{2}}{z_{h}^{2}},\nonumber \\
B_{B} & =-\ell_{\mathrm{AdS}}^{2}\left(\frac{1}{z_{B}^{2}}-\frac{1}{z_{h}^{2}}\right)dt_{B}\wedge dx_{B},\qquad\phi=\phi_{0},\qquad\beta=2\pi z_{h}.\label{eq:BTZ background}
\end{align}
The right exterior has $x_{B}\in\mathbb{R}$ and $0<z_{B}<z_{h}$.
The Kalb--Ramond potential is regular at the horizon, and its field
strength has the same orientation as in the vacuum construction: 
\begin{equation}
H=dB_{B}=\frac{2\ell_{\mathrm{AdS}}^{2}}{z_{B}^{3}}dt_{B}\wedge dx_{B}\wedge dz_{B},\qquad H_{t_{B}x_{B}z_{B}}=\frac{2\ell_{\mathrm{AdS}}^{2}}{z_{B}^{3}}.\label{eq:BTZ H}
\end{equation}
To simplify the calculation, we introduce the notation 
\begin{equation}
\begin{gathered}L_{A}=x_{2}-x_{1},\qquad x_{c}=\frac{x_{1}+x_{2}}{2},\qquad d=\frac{L_{A}}{2z_{h}},\qquad h_{B}=\sqrt{f_{B}},\\
T=\frac{t_{B}}{z_{h}},\qquad\xi=\frac{x_{B}-x_{c}}{z_{h}},\qquad p=\frac{x_{B}-x_{2}}{z_{h}}=\xi-d,
\end{gathered}
\label{eq:BTZ notation}
\end{equation}
where we use $z_{h}=\beta/\left(2\pi\right)$ as the thermal length
scale, so that $T$, $\xi$, $p$, and the interval half-width $d$
are dimensionless. The coordinate $\xi$ is centered at $x_{c}$,
making the symmetry of the RT surface explicit, while $p$ is measured
from $x_{2}$, the reference point for the F1 embeddings. Using $p=\xi-d$,
we express the string profiles in terms of $\xi$, the coordinate
used for the RT surface.

As in the vacuum case, we use the right endpoint $\left(0,x_{2},0\right)$
as the reference point and choose dimensionless worldsheet coordinates.
To construct the embedding, let $W=z_{h}/z_{B}$ and use 
\begin{equation}
\tau=Wh_{B}\sinh T,\qquad\sigma=-W\sinh p.\label{eq:BTZ gauge}
\end{equation}
Introduce the auxiliary combinations 
\begin{equation}
P=W\cosh p,\qquad Q=Wh_{B}\cosh T,\qquad P^{2}-\sigma^{2}=W^{2},\qquad Q^{2}-\tau^{2}=W^{2}-1.\label{eq:BTZ target combinations}
\end{equation}
These are functions of the target coordinates. Their differentials
give 
\begin{align}
-dP^{2}+d\sigma^{2} & =-dW^{2}+W^{2}dp^{2},\nonumber \\
dQ^{2}-d\tau^{2} & =\frac{W^{2}}{W^{2}-1}dW^{2}-\left(W^{2}-1\right)dT^{2}.
\end{align}
Based on these transformations, the BTZ line element pulled back to
the worldsheet therefore takes the form 
\begin{equation}
ds_{\mathrm{ws}}^{2}=\ell_{\mathrm{AdS}}^{2}\left[-d\tau^{2}+d\sigma^{2}-d\left(P-Q\right)d\left(P+Q\right)\right].\label{eq:BTZ pullback}
\end{equation}
We choose the branch $d\left(P-Q\right)=0$, for which the induced
metric is flat and the worldsheet metric constraints are automatic.
The embedding equations will be checked separately as vacuum case.
Writing the positive integration constant as $\eta$, we have 
\begin{equation}
P_{\eta}-Q_{\eta}=\eta,\qquad-P_{\eta}^{2}-\tau^{2}+\sigma^{2}+Q_{\eta}^{2}=-1.\label{eq:BTZ ambient constraints}
\end{equation}
Using the notation $\mathcal{D}=1+\sigma^{2}-\tau^{2}$, these two
relations yield 
\begin{equation}
P_{\eta}=\frac{\mathcal{D}+\eta^{2}}{2\eta},\qquad Q_{\eta}=P_{\eta}-\eta,\qquad W_{\eta}=\sqrt{P_{\eta}^{2}-\sigma^{2}}.\label{eq:BTZ auxiliary}
\end{equation}
The solution for the embedding is thus 
\begin{equation}
X^{t_{B}}\left(\tau,\sigma\right)=z_{h}\mathrm{arc}\tanh\frac{\tau}{Q_{\eta}},\qquad X^{x_{B}}\left(\tau,\sigma\right)=x_{2}-z_{h}\mathrm{arc}\tanh\frac{\sigma}{P_{\eta}},\qquad X^{z_{B}}\left(\tau,\sigma\right)=\frac{z_{h}}{W_{\eta}},\label{eq:BTZ embedding}
\end{equation}
where $Q_{\eta}>|\tau|$ and $P_{\eta}>|\sigma|$. In this domain
$W_{\eta}^{2}=1+Q_{\eta}^{2}-\tau^{2}>1$, so the embedding lies in
the right exterior. Since $dP_{\eta}=dQ_{\eta}$ at fixed $\eta$,
Eq. (\ref{eq:BTZ pullback}) gives 
\begin{equation}
ds_{\mathrm{ws}}^{2}=\ell_{\mathrm{AdS}}^{2}\left(-d\tau^{2}+d\sigma^{2}\right).\label{eq:TF1-induced}
\end{equation}
We now choose the same Weyl representative as in the vacuum calculation:
\begin{align}
\gamma_{mn} & =g_{\mathrm{ab}}\partial_{m}X^{\mathrm{a}}\partial_{n}X^{\mathrm{b}}=\ell_{\mathrm{AdS}}^{2}\mathrm{diag}\left(-1,1\right),\nonumber \\
\sqrt{-\det\gamma} & =\ell_{\mathrm{AdS}}^{2},\qquad\sqrt{-\gamma}\gamma^{mn}=\mathrm{diag}\left(-1,1\right).\label{eq:BTZ worldsheet metric}
\end{align}
The worldsheet metric constraints are then satisfied.

To identify the surface swept out by the embedding in target space,
we eliminate the worldsheet coordinates $\left(\tau,\sigma\right)$.
Define 
\begin{equation}
\Delta_{\beta}\equiv\cosh p-h_{B}\cosh T.
\end{equation}
By Eq. (\ref{eq:BTZ target combinations}), we have $W\Delta_{\beta}=P-Q$.
Since $P-Q=\eta$ on the chosen branch and $W=z_{h}/z_{B}$, the embedding
obeys
\begin{equation}
\Delta_{\beta}=\cosh\left(\frac{x_{B}-x_{2}}{z_{h}}\right)-\sqrt{1-\frac{z_{B}^{2}}{z_{h}^{2}}}\cosh\left(\frac{t_{B}}{z_{h}}\right)=\eta\frac{z_{B}}{z_{h}}.\label{eq:BTZ surface}
\end{equation}
For fixed $\eta>0$, this equation gives the target space image of
the string worldsheet in the right exterior $0<z_{B}<z_{h}$. The
worldsheet coordinates of a point on this surface are recovered from
Eq. (\ref{eq:BTZ gauge}).

We next verify the complete F1 equations, $\mathcal{E}^{\mathrm{a}}-H^{\mathrm{a}}{}_{\mathrm{bc}}\partial_{\tau}X^{\mathrm{b}}\partial_{\sigma}X^{\mathrm{c}}=0$,
with $\mathcal{E}^{\mathrm{a}}$ the metric contribution in conformal
gauge. Substituting Eq. (\ref{eq:BTZ embedding}) into the metric
and NS--NS terms separately gives
\begin{align}
\mathcal{E}^{t_{B}} & =H^{t_{B}}{}_{\mathrm{bc}}\partial_{\tau}X^{\mathrm{b}}\partial_{\sigma}X^{\mathrm{c}}=-\frac{2z_{B}^{2}\sinh T}{z_{h}h_{B}\Delta_{\beta}},\nonumber \\
\mathcal{E}^{x_{B}} & =H^{x_{B}}{}_{\mathrm{bc}}\partial_{\tau}X^{\mathrm{b}}\partial_{\sigma}X^{\mathrm{c}}=-\frac{2z_{B}^{2}\sinh p}{z_{h}\Delta_{\beta}},\nonumber \\
\mathcal{E}^{z_{B}} & =H^{z_{B}}{}_{\mathrm{bc}}\partial_{\tau}X^{\mathrm{b}}\partial_{\sigma}X^{\mathrm{c}}=\frac{2z_{B}h_{B}}{\Delta_{\beta}}\left(h_{B}\cosh p-\cosh T\right).\label{eq:BTZ eom check}
\end{align}
Note that $\mathcal{E}^{\mathrm{a}}$ is given by Eq. (\ref{eq:XI}).
The metric and NS--NS contributions therefore agree component by
component, so Eq. (\ref{eq:BTZ embedding}) describes on-shell F1
worldsheets.

To compare the string with the RT surface, we restrict the worldsheet
surface in Eq. (\ref{eq:BTZ surface}) to the $t_{B}=0$ slice. At
$t_{B}=0$, we use 
\begin{equation}
D_{\beta}\equiv\left.\Delta_{\beta}\right|_{t_{B}=0}=\cosh p-h_{B}.\label{eq:BTZ D}
\end{equation}
The spatial F1 profile is $D_{\beta}=\eta z_{B}/z_{h}$. We now compare
it with the RT surface \cite{Ryu:2006ef,Mintchev:2022fcp,Caggioli:2024uza}.
For a curve $z_{B}=z_{B}\left(x_{B}\right)$, its spatial length functional
and first integral are 
\begin{equation}
\mathcal{L}_{\gamma}=\ell_{\mathrm{AdS}}\int\frac{dx_{B}}{z_{B}}\sqrt{1+\frac{z_{B}^{\prime2}}{f_{B}}},\qquad\frac{1}{z_{B}\sqrt{1+z_{B}^{\prime2}/f_{B}}}=\frac{1}{z_{*}},\qquad z_{B}^{\prime}\equiv\frac{dz_{B}}{dx_{B}},\label{eq:BTZ geodesic integral}
\end{equation}
where $z_{*}$ is the turning point at $x_{B}=x_{c}$. Integrating
from $z_{B}$ to $z_{*}$ gives 
\begin{equation}
|x_{B}-x_{c}|=\int_{z_{B}}^{z_{*}}\frac{\widetilde{z}d\widetilde{z}}{\sqrt{\left(1-\widetilde{z}^{2}/z_{h}^{2}\right)\left(z_{*}^{2}-\widetilde{z}^{2}\right)}}=z_{h}\mathrm{arc}\cosh\frac{h_{B}}{\sqrt{1-z_{*}^{2}/z_{h}^{2}}},
\end{equation}
where $\widetilde{z}$ is a dummy radial integration variable, whereas
$z_{B}$ is the radial coordinate of the point on the surface. At
the ideal endpoints, $|x_{B}-x_{c}|=L_{A}/2$ and $h_{B}=1$. Hence
\begin{equation}
\gamma_{\beta,A}:\qquad\cosh\xi=h_{B}\cosh d,\qquad z_{*}=z_{h}\tanh d.\label{eq:BTZ RT}
\end{equation}

To check orthogonality, let $F_{\gamma}=\cosh\xi-h_{B}\cosh d$ and
$F_{\mathrm{F1}}=D_{\beta}-\eta z_{B}/z_{h}$ denote the two curve
equations. At their intersection the contraction of their normal covectors
in the spatial string-frame metric is 
\begin{align}
h^{ij}\partial_{i}F_{\gamma}\partial_{j}F_{\mathrm{F1}} & =\frac{z_{B}^{2}}{\ell_{\mathrm{AdS}}^{2}z_{h}^{2}}\left[\sinh\xi\sinh p+\cosh d\left(\frac{z_{B}^{2}}{z_{h}^{2}}-\eta\frac{z_{B}}{z_{h}}h_{B}\right)\right]\nonumber \\
 & =\frac{z_{B}^{2}}{\ell_{\mathrm{AdS}}^{2}z_{h}^{2}}\left[\cosh d-\cosh\left(\xi-p\right)\right]=0.\label{eq:BTZ orthogonality}
\end{align}
where we used $\eta z_{B}/z_{h}=\cosh p-h_{B}$, $h_{B}^{2}+z_{B}^{2}/z_{h}^{2}=1$,
and $h_{B}\cosh d=\cosh\xi$. The spatial F1 profiles therefore cross
the thermal RT surface orthogonally.

\subsection{The physical string current and thermal interval entropy}

\label{subsec:thermal-entropy}

To coarse-grain the discrete string current, we first determine the
string spacing with the thermal analogue of the vacuum regulator.
Keeping $x_{1}$ and $x_{2}$ fixed, we truncate the RT surface of
Eqs. (\ref{eq:BTZ geodesic integral}) to (\ref{eq:BTZ RT}) at $z_{B}=\varepsilon$,
with $0<\varepsilon<z_{*}$. We denote the truncated curve by $\gamma_{\beta,A,\varepsilon}$.
Eq. (\ref{eq:BTZ RT}) gives 
\begin{equation}
f_{\varepsilon}=1-\frac{\varepsilon^{2}}{z_{h}^{2}},\qquad\xi_{\varepsilon}=\mathrm{arc}\cosh\left(\sqrt{f_{\varepsilon}}\cosh d\right),\qquad a_{\beta,\varepsilon}=x_{c}-z_{h}\xi_{\varepsilon},\qquad b_{\beta,\varepsilon}=x_{c}+z_{h}\xi_{\varepsilon},\label{eq:BTZ regulator}
\end{equation}
with $A_{\beta,\varepsilon}=\left\{ z_{B}=\varepsilon,a_{\beta,\varepsilon}\le x_{B}\le b_{\beta,\varepsilon}\right\} $.
At $t_{B}=0$, the string passing through a cutoff point $x_{B}$
has integration constant 
\begin{equation}
\eta\left(x_{B}\right)=\frac{z_{h}}{\varepsilon}\left[\cosh\frac{x_{B}-x_{2}}{z_{h}}-\sqrt{f_{\varepsilon}}\right].\label{eq:BTZ cutoff label}
\end{equation}
Since $\xi_{\varepsilon}<d$, one has $p<0$ on the regulated interval
and $d\eta/dx_{B}=\sinh p/\varepsilon<0$. The minimum and maximum
therefore occur at $b_{\beta,\varepsilon}$ and $a_{\beta,\varepsilon}$,
respectively: 
\begin{equation}
\eta_{\min}=\frac{z_{h}}{\varepsilon}\left[\cosh\left(d-\xi_{\varepsilon}\right)-\sqrt{f_{\varepsilon}}\right],\qquad\eta_{\max}=\frac{z_{h}}{\varepsilon}\left[\cosh\left(d+\xi_{\varepsilon}\right)-\sqrt{f_{\varepsilon}}\right].\label{eq:BTZ label endpoints}
\end{equation}
Using $\cosh\xi_{\varepsilon}=\sqrt{f_{\varepsilon}}\cosh d$, these
can be rewritten as
\begin{equation}
\eta_{\min,\max}=\frac{z_{h}\sinh d}{\varepsilon}\left(\sqrt{f_{\varepsilon}}\sinh d\mp\sinh\xi_{\varepsilon}\right),\qquad f_{\varepsilon}\sinh^{2}d-\sinh^{2}\xi_{\varepsilon}=\frac{\varepsilon^{2}}{z_{h}^{2}}.\label{eq:BTZ cutoff identities}
\end{equation}
Each value of the integration constant $\eta$ specifies one F1 profile.
The strings that meet the regulated interval $A_{\beta,\varepsilon}$
have $\eta\in\left[\eta_{\min},\eta_{\max}\right]$. We define the
half-width of this range in $\log\eta$ by

\begin{equation}
U_{\beta,\varepsilon}\equiv\frac{1}{2}\log\frac{\eta_{\max}}{\eta_{\min}}=\log\left[\frac{z_{h}}{\varepsilon}\left(\sqrt{f_{\varepsilon}}\sinh d+\sinh\xi_{\varepsilon}\right)\right]=\mathrm{arc}\cosh\left(\frac{z_{h}}{\varepsilon}\sqrt{f_{\varepsilon}}\sinh d\right).\label{eq:BTZU}
\end{equation}
Since $\eta_{\min}\eta_{\max}=\sinh^{2}d$, the endpoints can be written
as 
\begin{equation}
\eta_{\min}=\sinh de^{-U_{\beta,\varepsilon}},\qquad\eta_{\max}=\sinh de^{U_{\beta,\varepsilon}}.\label{eq:BTZ label range}
\end{equation}

\begin{figure}[h]
\begin{centering}
\includegraphics[scale=1.4]{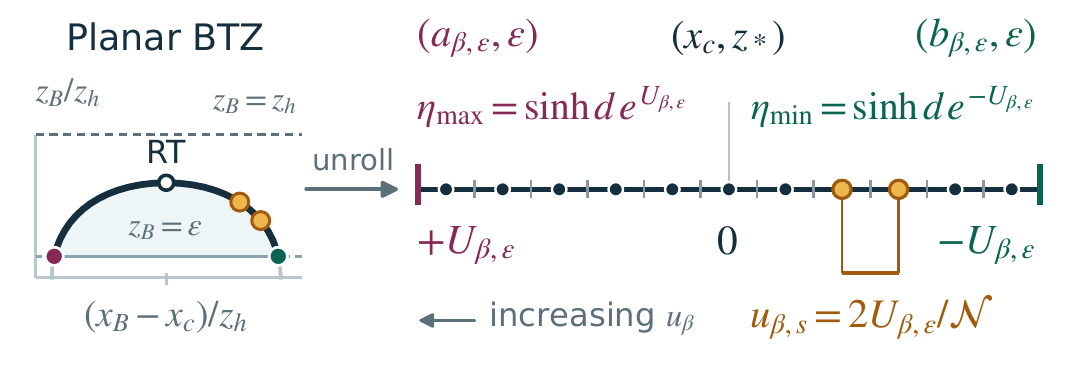}
\par\end{centering}
\caption{\label{fig:thermal ruler}The regulated thermal RT surface in planar
BTZ at $t_{B}=0$ is shown on the left and unfolded onto the arclength
ruler on the right. The left and right cutoff endpoints map to $+U_{\beta,\varepsilon}$
and $-U_{\beta,\varepsilon}$, corresponding to $\eta_{\max}$ and
$\eta_{\min}$, respectively. The open circle marks the turning point
$\left(x_{c},z_{*}\right)$. The dots on the ruler mark F1 crossings
at the cell midpoints. The gold dots identify the same pair of adjacent
crossings in both panels, with proper separation $\ell_{\mathrm{AdS}}u_{\beta,s}$
along the surface.}
\end{figure}

We now explain why the ensemble is uniform in $\log\eta$. Introduce
the dimensionless notation
\begin{equation}
u_{\beta}=\log\frac{\eta}{\sinh d},\qquad-U_{\beta,\varepsilon}\le u_{\beta}\le U_{\beta,\varepsilon}.\label{eq:BTZ log label}
\end{equation}
On the RT surface, $D_{\beta}=-\tanh d\sinh p$. Solving the F1 and
RT curve equations together then gives 
\begin{equation}
\left(\frac{\eta}{\sinh d}\right)^{2}=\frac{\tanh d-\tanh\xi}{\tanh d+\tanh\xi},\qquad\tanh\xi=-\tanh d\tanh u_{\beta},\qquad\frac{z_{B}}{z_{h}}=\frac{\sinh d}{\sqrt{\cosh^{2}u_{\beta}+\sinh^{2}d}}.\label{eq:BTZ RT label}
\end{equation}
In particular, differentiating $x_{B}=x_{c}+z_{h}\xi$ along this
curve gives $|dx_{B}/du_{\beta}|=z_{B}^{2}/z_{*}$. Combining this
relation with Eq. (\ref{eq:BTZ geodesic integral}) gives 
\begin{equation}
dA_{S}=\frac{\ell_{\mathrm{AdS}}}{z_{B}}\sqrt{1+\frac{z_{B}^{\prime2}}{f_{B}}}|dx_{B}|=\frac{\ell_{\mathrm{AdS}}z_{*}}{z_{B}^{2}}|dx_{B}|=\ell_{\mathrm{AdS}}|du_{\beta}|.\label{eq:TF1-RT-length-element}
\end{equation}
Therefore, the equal spacing in $u_{\beta}$ gives equal proper spacing
on the RT surface, just as in the vacuum case. The bundle also covers
the regulated homology region: at fixed $z_{B}$, $u_{\beta}$ decreases
with $x_{B}$ and its values between the two branches of the RT surface
lie between the labels on those branches. We then divide the range
into $\mathcal{N}$ equal cells and put one string at each midpoint:
\begin{align}
u_{\beta,s} & =\frac{2U_{\beta,\varepsilon}}{\mathcal{N}},\qquad u_{\beta,k}=-U_{\beta,\varepsilon}+\left(k+\frac{1}{2}\right)u_{\beta,s},\nonumber \\
\eta_{k} & =\sinh de^{u_{\beta,k}}=\eta_{\min}\exp\left[\left(k+\frac{1}{2}\right)u_{\beta,s}\right],\qquad k=0,\ldots,\mathcal{N}-1.\label{eq:BTZ discrete population}
\end{align}
 The adjacent strings satisfy $\log\left(\eta_{k+1}/\eta_{k}\right)=u_{\beta,s}$.
Fig. (\ref{fig:thermal ruler}) shows the corresponding midpoint sampling
for the thermal interval: the left and right cutoff endpoints carry
the labels $\eta_{\max}$ and $\eta_{\min}$, respectively, and the
F1 crossings are equally spaced in $u_{\beta}$.

We now return to the normalized current of $\mathcal{N}$ equally
oriented strings. Writing $x=(t_{B},x_{B},z_{B})$ for the observation
event, we have 
\begin{equation}
j^{\mathrm{ab}}\left(x\right)=\frac{1}{\sqrt{-g}}\sum_{k=0}^{\mathcal{N}-1}\int d\tau d\sigma\left(\partial_{\tau}X_{k}^{\mathrm{a}}\partial_{\sigma}X_{k}^{\mathrm{b}}-\partial_{\sigma}X_{k}^{\mathrm{a}}\partial_{\tau}X_{k}^{\mathrm{b}}\right)\delta^{\left(3\right)}\left(x-X_{k}\right).\label{eq:BTZ source current}
\end{equation}
where $X_{k}$ denotes the embedding in Eq. (\ref{eq:BTZ embedding})
with $\eta=\eta_{k}$. Before setting $t_{B}=0$, we invert the embedding
family using Eqs. (\ref{eq:BTZ gauge}) and (\ref{eq:BTZ surface}):
\begin{equation}
\tau_{*}=\frac{z_{h}h_{B}}{z_{B}}\sinh T,\qquad\sigma_{*}=-\frac{z_{h}}{z_{B}}\sinh p,\qquad u_{\beta}=\log\frac{z_{h}\Delta_{\beta}}{z_{B}\sinh d}.\label{eq:BTZ inverse map}
\end{equation}
This inverse is defined for $\Delta_{\beta}>0$ in the right exterior.
On the $k$th worldsheet, the inverse map gives $u_{\beta}=u_{\beta,k}$,
while $\tau_{*}$ and $\sigma_{*}$ are the worldsheet coordinates
of the observation event.

Allowing $u_{\beta}$ to vary across the family, differentiation of
the inverse map in $\left(T,p,W\right)$, with $W=z_{h}/z_{B}$, gives
\begin{align}
d\tau & =Wh_{B}\cosh TdT+\frac{\sinh T}{h_{B}}dW,\nonumber \\
d\sigma & =-W\cosh pdp-\sinh pdW,\nonumber \\
du_{\beta} & =\frac{-h_{B}\sinh TdT+\sinh pdp}{\Delta_{\beta}}+\frac{\cosh p-\cosh T/h_{B}}{W\Delta_{\beta}}dW.
\end{align}
Using $\cosh^{2}p-\sinh^{2}p=\cosh^{2}T-\sinh^{2}T=1$, the wedge
product reduces to 
\begin{equation}
d\tau\wedge d\sigma\wedge du_{\beta}=WdT\wedge dp\wedge dW=-\frac{1}{z_{B}^{3}}dt_{B}\wedge dx_{B}\wedge dz_{B}.\label{eq:BTZ jacobian}
\end{equation}
The absolute Jacobian is therefore $z_{B}^{3}$, and 
\begin{equation}
\delta^{\left(3\right)}\left(x-X_{k}\right)=\frac{1}{z_{B}^{3}}\delta\left(\tau-\tau_{*}\right)\delta\left(\sigma-\sigma_{*}\right)\delta\left(u_{\beta}-u_{\beta,k}\right).\label{eq:BTZ delta transformation}
\end{equation}
The first two delta functions perform the two worldsheet integrations.
On the support of the remaining delta function, one may set $u_{\beta,k}=u_{\beta}\left(x\right)$
in the derivative products, giving 
\begin{equation}
j^{\mathrm{ab}}=\frac{1}{\sqrt{-g}z_{B}^{3}}\left(\partial_{\tau}X^{\mathrm{a}}\partial_{\sigma}X^{\mathrm{b}}-\partial_{\sigma}X^{\mathrm{a}}\partial_{\tau}X^{\mathrm{b}}\right)_{*}\sum_{k=0}^{\mathcal{N}-1}\delta\left(u_{\beta}-u_{\beta,k}\right),\label{eq:BTZ current integrated}
\end{equation}
where the subscript $*$ also indicates that the worldsheet derivatives
are taken at fixed $u_{\beta}$ and then evaluated at $\left(\tau_{*},\sigma_{*},u_{\beta}\left(x\right)\right)$.

Substitution of the embedding into the three independent derivative
products gives 
\begin{align}
\big(\partial_{\tau}X^{t_{B}}\partial_{\sigma}X^{x_{B}}-\partial_{\sigma}X^{t_{B}}\partial_{\tau}X^{x_{B}}\big)_{*} & =\frac{z_{B}^{2}\left(h_{B}\cosh p-\cosh T\right)}{h_{B}\Delta_{\beta}},\nonumber \\
\big(\partial_{\tau}X^{t_{B}}\partial_{\sigma}X^{z_{B}}-\partial_{\sigma}X^{t_{B}}\partial_{\tau}X^{z_{B}}\big)_{*} & =\frac{z_{B}^{3}\sinh p}{z_{h}\Delta_{\beta}},\nonumber \\
\big(\partial_{\tau}X^{x_{B}}\partial_{\sigma}X^{z_{B}}-\partial_{\sigma}X^{x_{B}}\partial_{\tau}X^{z_{B}}\big)_{*} & =\frac{z_{B}^{3}h_{B}\sinh T}{z_{h}\Delta_{\beta}}.\label{eq:BTZ bivectors}
\end{align}
 With $\sqrt{-g}=\ell_{\mathrm{AdS}}^{3}/z_{B}^{3}$, the discrete
current is 
\begin{align}
j^{0x_{B}} & =\frac{z_{B}^{2}\left(h_{B}\cosh p-\cosh T\right)}{\ell_{\mathrm{AdS}}^{3}h_{B}\Delta_{\beta}}\sum_{k=0}^{\mathcal{N}-1}\delta\left(u_{\beta}-u_{\beta,k}\right),\nonumber \\
j^{0z_{B}} & =\frac{z_{B}^{3}\sinh p}{\ell_{\mathrm{AdS}}^{3}z_{h}\Delta_{\beta}}\sum_{k=0}^{\mathcal{N}-1}\delta\left(u_{\beta}-u_{\beta,k}\right),\nonumber \\
j^{x_{B}z_{B}} & =\frac{z_{B}^{3}h_{B}\sinh T}{\ell_{\mathrm{AdS}}^{3}z_{h}\Delta_{\beta}}\sum_{k=0}^{\mathcal{N}-1}\delta\left(u_{\beta}-u_{\beta,k}\right).\label{eq:BTZ discrete current}
\end{align}

Next, we coarse-grain the discrete sum. Since $u_{\beta}$ is nonperiodic,
the delta functions are ordinary delta functions. We write
\begin{align}
\sum_{k=0}^{\mathcal{N}-1}\delta\left(u_{\beta}-u_{\beta,k}\right) & =\frac{1}{u_{\beta,s}}\left[u_{\beta,s}\sum_{k=0}^{\mathcal{N}-1}\delta\left(u_{\beta}-u_{\beta,k}\right)\right]\nonumber \\
 & =\frac{\mathcal{N}}{2U_{\beta,\varepsilon}}\left[u_{\beta,s}\sum_{k=0}^{\mathcal{N}-1}\delta\left(u_{\beta}+U_{\beta,\varepsilon}-\left(k+\tfrac{1}{2}\right)u_{\beta,s}\right)\right].\label{eq:BTZ delta comb}
\end{align}
 For a smooth test function $f$, the expression in brackets acts
as the midpoint Riemann sum 
\begin{equation}
u_{\beta,s}\sum_{k=0}^{\mathcal{N}-1}f\left(u_{\beta,k}\right)\longrightarrow\int_{-U_{\beta,\varepsilon}}^{U_{\beta,\varepsilon}}f\left(u_{\beta}\right)du_{\beta},
\end{equation}
as $\mathcal{N}\to\infty$ at fixed $U_{\beta,\varepsilon}$. Its
weak limit is one inside the range of string labels and zero outside.
Hence 
\begin{equation}
\sum_{k=0}^{\mathcal{N}-1}\delta\left(u_{\beta}-u_{\beta,k}\right)\xrightarrow{\mathrm{coarse\text{-}graining}}\frac{\mathcal{N}}{2U_{\beta,\varepsilon}},\qquad-U_{\beta,\varepsilon}<u_{\beta}<U_{\beta,\varepsilon}.\label{eq:BTZ coarse graining}
\end{equation}
Substituting this density into the discrete current gives the smooth
current at general target-space time: 
\begin{equation}
j^{0x_{B}}=\frac{\mathcal{N}z_{B}^{2}\left(h_{B}\cosh p-\cosh T\right)}{2U_{\beta,\varepsilon}\ell_{\mathrm{AdS}}^{3}h_{B}\Delta_{\beta}},\qquad j^{0z_{B}}=\frac{\mathcal{N}z_{B}^{3}\sinh p}{2U_{\beta,\varepsilon}\ell_{\mathrm{AdS}}^{3}z_{h}\Delta_{\beta}},\qquad j^{x_{B}z_{B}}=\frac{\mathcal{N}z_{B}^{3}h_{B}\sinh T}{2U_{\beta,\varepsilon}\ell_{\mathrm{AdS}}^{3}z_{h}\Delta_{\beta}},\label{eq:BTZ smooth current}
\end{equation}
where $\Delta_{\beta}>0$ and $|u_{\beta}|<U_{\beta,\varepsilon}$.
The densitized components $\sqrt{-g}\left(j^{0x_{B}},j^{0z_{B}},j^{x_{B}z_{B}}\right)$
are proportional to $\left(-\partial_{z_{B}}u_{\beta},\partial_{x_{B}}u_{\beta},-\partial_{t_{B}}u_{\beta}\right)$,
so the spacetime current is conserved. 

We now project onto the constant-$t_{B}$ slices. The lapse, spatial
metric, and spatial volume density are 
\begin{equation}
N_{\mathrm{ADM}}=\frac{\ell_{\mathrm{AdS}}h_{B}}{z_{B}},\qquad h_{ij}dx^{i}dx^{j}=\frac{\ell_{\mathrm{AdS}}^{2}}{z_{B}^{2}}\left(dx_{B}^{2}+\frac{dz_{B}^{2}}{f_{B}}\right),\qquad\sqrt{h}=\frac{\ell_{\mathrm{AdS}}^{2}}{z_{B}^{2}h_{B}}.\label{eq:BTZ ADM}
\end{equation}
The future-directed unit normal has covariant components $n_{\mu}=\left(-N_{\mathrm{ADM}},0,0\right)$.
Thus $q^{\mu}=-n_{\nu}j^{\nu\mu}$, $q^{0}=0$, and $q^{i}=N_{\mathrm{ADM}}j^{0i}$
for $i=x_{B},z_{B}$, which yields 

\begin{equation}
q^{x_{B}}=\frac{\mathcal{N}z_{B}\left(h_{B}\cosh p-\cosh T\right)}{2U_{\beta,\varepsilon}\ell_{\mathrm{AdS}}^{2}\Delta_{\beta}},\qquad q^{z_{B}}=\frac{\mathcal{N}z_{B}^{2}h_{B}\sinh p}{2U_{\beta,\varepsilon}\ell_{\mathrm{AdS}}^{2}z_{h}\Delta_{\beta}}.\label{eq:BTZ general q}
\end{equation}
 At fixed $t_{B}$ the projected current is conserved: 
\begin{equation}
\partial_{x_{B}}\left(\sqrt{h}q^{x_{B}}\right)=-\partial_{z_{B}}\left(\sqrt{h}q^{z_{B}}\right)=\frac{\mathcal{N}z_{B}\cosh T\sinh p}{2U_{\beta,\varepsilon}z_{h}^{3}h_{B}\Delta_{\beta}^{2}},\qquad\nabla_{i}q^{i}=0.\label{eq:BTZ general conservation}
\end{equation}
The current is tangent to the level sets of $u_{\beta}$ on each slice.
Its string-frame norm is 
\begin{align}
|q|^{2} & =\frac{\ell_{\mathrm{AdS}}^{2}}{z_{B}^{2}}\left[\left(q^{x_{B}}\right)^{2}+\frac{\left(q^{z_{B}}\right)^{2}}{f_{B}}\right]\nonumber \\
 & =\left(\frac{\mathcal{N}}{2\ell_{\mathrm{AdS}}U_{\beta,\varepsilon}}\right)^{2}\frac{\left(h_{B}\cosh p-\cosh T\right)^{2}+\left(z_{B}^{2}/z_{h}^{2}\right)\sinh^{2}p}{\Delta_{\beta}^{2}}\nonumber \\
 & =\left(\frac{\mathcal{N}}{2\ell_{\mathrm{AdS}}U_{\beta,\varepsilon}}\right)^{2}\left(1+\frac{z_{B}^{2}\sinh^{2}T}{z_{h}^{2}\Delta_{\beta}^{2}}\right).\label{eq:BTZ general norm}
\end{align}
 The last equality follows from $h_{B}^{2}+z_{B}^{2}/z_{h}^{2}=1$.

\begin{figure}[h]
\begin{centering}
\includegraphics[scale=0.18]{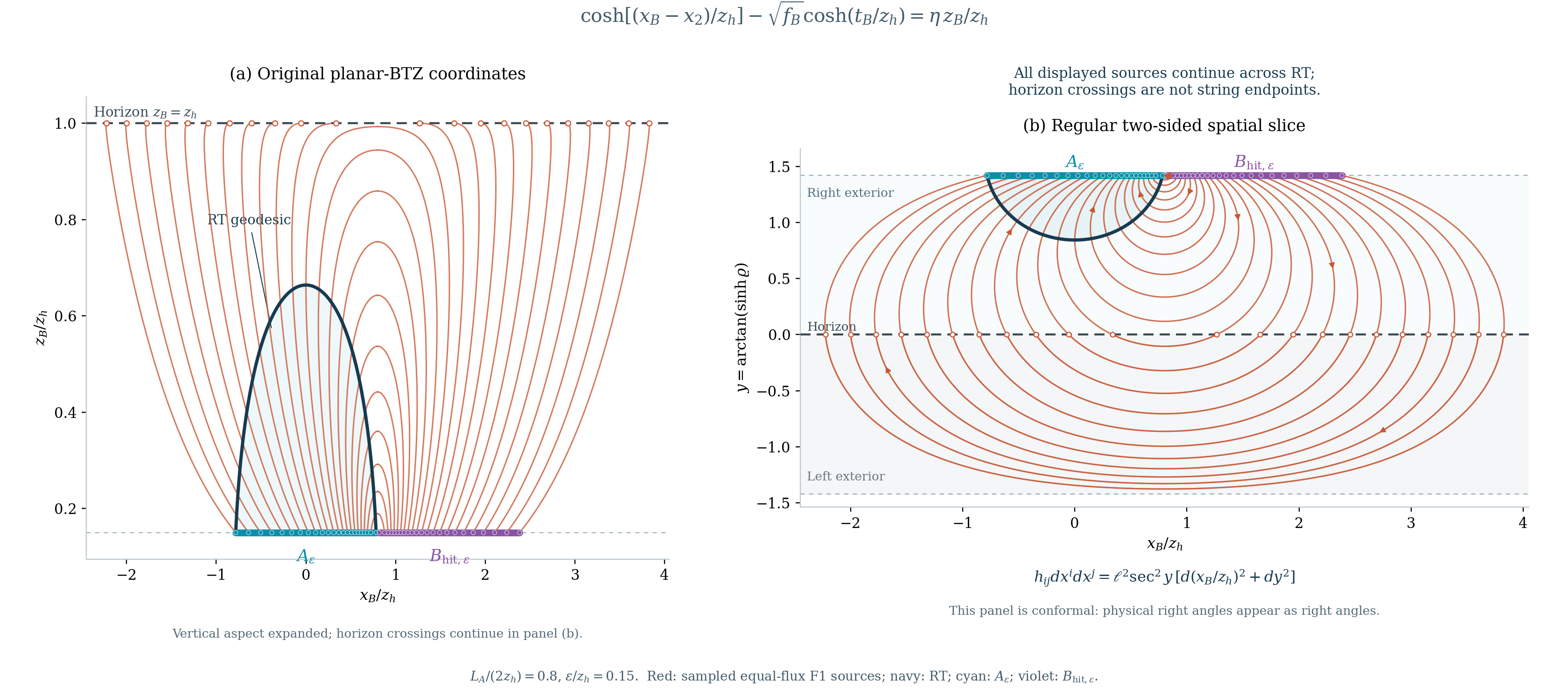}
\par\end{centering}
\caption{\label{fig:thermal}The red curves are the $t_{B}=0$ profiles $\cosh p-h_{B}=\eta z_{B}/z_{h}$,
sampled uniformly in $u_{\beta}=\log\left(\eta/\sinh d\right)$. They
cross the thermal RT surface $\gamma_{\beta,A}$ (navy) orthogonally
and at equal proper intervals along it. Panel (a) shows the right
exterior in the original BTZ coordinates, with an expanded vertical
scale. Panel (b) shows the same profiles on a regular two-sided conformal
slice, where physical angles are preserved and strings that reach
the horizon continue across the bifurcation line. Cyan marks $A_{\beta,\varepsilon}$,
and violet marks the subset of its regulated complement reached by
this bundle. The plot uses $L_{A}/\left(2z_{h}\right)=0.8$ and $\varepsilon/z_{h}=0.15$.}
\end{figure}

The ADM projection defines a spatial current on each slice, with $t_{B}$
labeling the slice. Since $A$ is defined at $t_{B}=0$ and its RT
surface lies on the same slice, we now set $t_{B}=0$ to compute the
entropy flux. On this slice $\Delta_{\beta}=D_{\beta}$, $j^{x_{B}z_{B}}=0$,
and 

\begin{equation}
q^{x_{B}}=\frac{\mathcal{N}z_{B}\left(h_{B}\cosh p-1\right)}{2U_{\beta,\varepsilon}\ell_{\mathrm{AdS}}^{2}D_{\beta}},\qquad q^{z_{B}}=\frac{\mathcal{N}z_{B}^{2}h_{B}\sinh p}{2U_{\beta,\varepsilon}\ell_{\mathrm{AdS}}^{2}z_{h}D_{\beta}},\label{eq:BTZ time zero q}
\end{equation}
with $|q|=\frac{\mathcal{N}}{2\ell_{\mathrm{AdS}}U_{\beta,\varepsilon}}$.
Fig. (\ref{fig:thermal}) shows representative profiles of the regulated
F1 bundle on the $t_{B}=0$ slice in planar BTZ and their crossings
of the thermal RT surface. 
 The outward unit normal and proper line element on the cutoff are
\begin{equation}
n_{\mathrm{cut}}=-\frac{\varepsilon\sqrt{f_{\varepsilon}}}{\ell_{\mathrm{AdS}}}\partial_{z_{B}},\qquad dA_{S}=\frac{\ell_{\mathrm{AdS}}}{\varepsilon}dx_{B}.\label{eq:BTZ cutoff normal}
\end{equation}
The charge flux through the full regulated interval is 
\begin{align}
\Phi_{\beta,A}\left[q\right] & \equiv\int_{A_{\beta,\varepsilon}}q\cdot n_{\mathrm{cut}}dA_{S}\nonumber \\
 & =-\frac{\mathcal{N}}{2U_{\beta,\varepsilon}z_{h}}\int_{a_{\beta,\varepsilon}}^{b_{\beta,\varepsilon}}\frac{\sinh\left[\left(x_{B}-x_{2}\right)/z_{h}\right]}{\cosh\left[\left(x_{B}-x_{2}\right)/z_{h}\right]-\sqrt{f_{\varepsilon}}}dx_{B}\nonumber \\
 & =\frac{\mathcal{N}}{2U_{\beta,\varepsilon}}\log\frac{\eta_{\max}}{\eta_{\min}}=\mathcal{N},\label{eq:BTZ charge flux}
\end{align}
where the last equality uses Eq. (\ref{eq:BTZU}).

Following the vacuum calculation, define the unit-normalized spatial
current inside the bundle by 

\begin{equation}
\widehat{q}^{i}=\frac{q^{i}}{|q|},\qquad\widehat{q}=\frac{z_{B}}{\ell_{\mathrm{AdS}}D_{\beta}}\left[\left(h_{B}\cosh p-1\right)\partial_{x_{B}}+\frac{z_{B}}{z_{h}}h_{B}\sinh p\partial_{z_{B}}\right].\label{eq:BTZ unit current}
\end{equation}
 Since $|q|$ is constant on this slice inside the bundle, $|\widehat{q}|=1$
and $\nabla_{i}\widehat{q}^{i}=\nabla_{i}q^{i}/|q|=0$. The resulting
flux is 
\begin{equation}
\Phi_{\beta,A}\left[\widehat{q}\right]=\frac{\Phi_{\beta,A}\left[q\right]}{|q|}=2\ell_{\mathrm{AdS}}U_{\beta,\varepsilon}.\label{eq:BTZ unit flux}
\end{equation}
As in the vacuum case, we use the string-frame entropy capacity from
the off-shell sphere and replica construction, 
\begin{equation}
\rho_{S}=\frac{e^{-2\phi_{0}}}{4G_{N}^{\left(3\right)}}.\label{eq:BTZ capacity}
\end{equation}
The constant dilaton makes $\rho_{S}$ constant. Combining this capacity
with the unit current gives the physical entropy flow 
\begin{align}
v_{\mathrm{phys}}^{i} & \equiv\rho_{S}\widehat{q}^{i}=\rho_{S}\frac{q^{i}}{|q|},\nonumber \\
v_{\mathrm{phys}} & =\frac{\rho_{S}z_{B}}{\ell_{\mathrm{AdS}}D_{\beta}}\left[\left(h_{B}\cosh p-1\right)\partial_{x_{B}}+\frac{z_{B}}{z_{h}}h_{B}\sinh p\partial_{z_{B}}\right].\label{eq:BTZ entropy flow}
\end{align}
It satisfies $\nabla_{i}v_{\mathrm{phys}}^{i}=0$ and $|v_{\mathrm{phys}}|=\rho_{S}$
inside the string bundle. Its flux is 
\begin{equation}
\Phi_{\beta,A}\equiv\Phi_{\beta,A}\left[v_{\mathrm{phys}}\right]=\rho_{S}\Phi_{\beta,A}\left[\widehat{q}\right]=2\rho_{S}\ell_{\mathrm{AdS}}U_{\beta,\varepsilon}.\label{eq:BTZ entropy flux}
\end{equation}
The physical F1 current thus realizes the maximal entropy flux, and
the entropy with the geometric regulator is 
\begin{equation}
S_{A}^{\left(0\right)}=\Phi_{\beta,A}=2\rho_{S}\ell_{\mathrm{AdS}}U_{\beta,\varepsilon}=2\rho_{S}\ell_{\mathrm{AdS}}\mathrm{arc}\cosh\left[\frac{z_{h}}{\varepsilon}\sqrt{f_{\varepsilon}}\sinh\frac{L_{A}}{2z_{h}}\right].\label{eq:BTZ exact entropy}
\end{equation}
Using central charge $c=6\rho_{S}\ell_{\mathrm{AdS}}=3\ell_{\mathrm{AdS}}e^{-2\phi_{0}}/\left(2G_{N}^{\left(3\right)}\right)$
and $\beta=2\pi z_{h}$, we obtain, at fixed $L_{A}$ and $\beta$, 

\begin{equation}
S_{A}^{\left(0\right)}=2\rho_{S}\ell_{\mathrm{AdS}}\log\left[\frac{2z_{h}}{\varepsilon}\sinh\frac{L_{A}}{2z_{h}}\right]+o\left(1\right)=\frac{c}{3}\log\left[\frac{\beta}{\pi\varepsilon}\sinh\frac{\pi L_{A}}{\beta}\right]+o\left(1\right),\label{eq:BTZ CFT entropy}
\end{equation}
with the limit $\varepsilon\to0$. The finite-cutoff terms refer to
the regulator in Eq. (\ref{eq:BTZ regulator}). Therefore, the continuum
result reproduces the well-known thermal interval entanglement entropy
on an infinite line \cite{Calabrese:2004eu}, obtained here from physical
F1 sources. 

Finally, to verify maximality, choose the RT unit normal pointing
toward the interval homology region: 
\begin{equation}
n_{\beta,A}=-\frac{z_{B}}{\ell_{\mathrm{AdS}}\sinh d}\left(\sinh\xi\partial_{x_{B}}+\frac{z_{B}}{z_{h}}h_{B}\cosh d\partial_{z_{B}}\right).\label{eq:BTZ RT normal}
\end{equation}
On $\gamma_{\beta,A}$, we have $D_{\beta}=-\tanh d\sinh p$ and $h_{B}\cosh p-1=\sinh\xi\sinh p/\cosh d$.
Substitution into Eq. (\ref{eq:BTZ entropy flow}) gives 
\begin{equation}
\left.v_{\mathrm{phys}}\right|_{\gamma_{\beta,A,\varepsilon}}=\rho_{S}n_{\beta,A}.\label{eq:BTZ RT saturation}
\end{equation}
The flow therefore saturates the capacity normally on the entire regulated
RT surface. Eq. (\ref{eq:TF1-RT-length-element}) gives $\mathrm{Length}\left(\gamma_{\beta,A,\varepsilon}\right)=2\ell_{\mathrm{AdS}}U_{\beta,\varepsilon}$.
For any admissible flow $w^{i}$ on the same regulated homology region,
with $\nabla_{i}w^{i}=0$ and $|w|\le\rho_{S}$, flux conservation
and the weighted flux bound \cite{Headrick:2017ucz,Wu:2026pgc} of
the off-shell construction give 
\begin{align}
\Phi_{\beta,A}\left[w\right] & =\int_{\gamma_{\beta,A,\varepsilon}}w\cdot n_{\beta,A}dA_{S}\le\int_{\gamma_{\beta,A,\varepsilon}}|w|dA_{S}\nonumber \\
 & \le\rho_{S}\mathrm{Length}\left(\gamma_{\beta,A,\varepsilon}\right)=2\rho_{S}\ell_{\mathrm{AdS}}U_{\beta,\varepsilon}=\Phi_{\beta,A}\left[v_{\mathrm{phys}}\right].\label{eq:BTZ maximality}
\end{align}

\subsection{Vacuum limit and thermal extensivity}

We finally check that the thermal construction reduces to the vacuum
worldsheet and current, and that its long-interval flux reproduces
the horizon entropy density.

For the vacuum limit, take $x_{c}=0$, $L_{A}=2R$, and let $z_{h}\to\infty$
with $z_{0}=z_{h}\eta$ and the cutoff $\varepsilon$ fixed. At fixed
$\left(\tau,\sigma\right)$ with $\mathcal{D}>0$, 
\begin{equation}
P_{\eta},\ Q_{\eta},\ W_{\eta}=\frac{z_{h}\mathcal{D}}{2z_{0}}+O\left(z_{h}^{-1}\right).
\end{equation}
The thermal embedding consequently tends to 
\begin{equation}
\left(X^{t_{B}},X^{x_{B}},X^{z_{B}}\right)\longrightarrow\left(\frac{2z_{0}\tau}{\mathcal{D}},R-\frac{2z_{0}\sigma}{\mathcal{D}},\frac{2z_{0}}{\mathcal{D}}\right),\label{eq:BTZ embedding vacuum limit}
\end{equation}
which is the vacuum embedding after identifying $\left(t_{B},x_{B},z_{B}\right)$
with $\left(t,x,z\right)$. At fixed bulk coordinates, the combinations
in the current obey 

\begin{equation}
\Delta_{\beta}=\frac{\Delta_{A}}{2z_{h}^{2}}+O\left(z_{h}^{-4}\right),\qquad\Delta_{A}=\left(x_{B}-R\right)^{2}+z_{B}^{2}-t_{B}^{2},
\end{equation}
and

\begin{equation}
h_{B}\cosh p-\cosh T=\frac{(x_{B}-R)^{2}-z_{B}^{2}-t_{B}^{2}}{2z_{h}^{2}}+O\left(z_{h}^{-4}\right).
\end{equation}
The label $u_{\beta}=\log\left[z_{h}\Delta_{\beta}/\left(z_{B}\sinh d\right)\right]$
tends to $u=\log\left[\Delta_{A}/\left(2Rz_{B}\right)\right]$. The
regulated half-width satisfies $U_{\beta,\varepsilon}\to\mathrm{arc}\cosh\left(R/\varepsilon\right)=U_{\varepsilon}$,
so the uniform distribution over $u_{\beta}$ reduces to the vacuum
distribution over $u$. Inside the string bundle, these expansions,
together with $\sinh p=p+O\left(p^{3}\right)$ and $\sinh T=T+O\left(T^{3}\right)$,
give 
\begin{equation}
\begin{pmatrix}j^{0x_{B}}\\
j^{0z_{B}}\\
j^{x_{B}z_{B}}
\end{pmatrix}\longrightarrow\frac{\mathcal{N}}{2U_{\varepsilon}\ell_{\mathrm{AdS}}^{3}\Delta_{A}}\begin{pmatrix}z_{B}^{2}\left[\left(x_{B}-R\right)^{2}-z_{B}^{2}-t_{B}^{2}\right]\\
2z_{B}^{3}\left(x_{B}-R\right)\\
2t_{B}z_{B}^{3}
\end{pmatrix}.\label{eq:BTZ current vacuum limit}
\end{equation}
Thus the full spacetime current, including its spatial component at
nonzero time, tends to the vacuum current. The regulated entropy flux
likewise tends to $2\rho_{S}\ell_{\mathrm{AdS}}U_{\varepsilon}$.

To check thermal extensivity, we ask how much entropy is added when
the interval is enlarged at fixed inverse temperature $\beta=2\pi z_{h}$.
The $L_{A}$-dependent term in the continuum entropy is $\left(c/3\right)\log\sinh d$,
with $d=L_{A}/\left(2z_{h}\right)=\pi L_{A}/\beta$. Its slope is
\begin{equation}
\left.\frac{\partial S_{A}^{\left(0\right)}}{\partial L_{A}}\right|_{\beta}=\frac{c}{3}\frac{\cosh d}{\sinh d}\frac{1}{2z_{h}}=\frac{c}{6z_{h}}\coth d.
\end{equation}
 For $L_{A}/\beta\to\infty$, we have $d\to\infty$ and $\coth d\to1$.
Thus, using $c=6\rho_{S}\ell_{\mathrm{AdS}}$, we obtain 
\begin{equation}
\lim_{L_{A}/\beta\to\infty}\left.\frac{\partial S_{A}^{\left(0\right)}}{\partial L_{A}}\right|_{\beta}=\frac{c}{6z_{h}}=\frac{\pi c}{3\beta}=\frac{\rho_{S}\ell_{\mathrm{AdS}}}{z_{h}}.\label{eq:f1-thermal-extensivity}
\end{equation}
The entropy therefore grows linearly at large $L_{A}$, with this
limiting slope giving the entropy added per unit length. On the BTZ
horizon, $\mathrm{d}A_{S}=\left(\ell_{\mathrm{AdS}}/z_{h}\right)\mathrm{d}x_{B}$.
Multiplying the proper length per unit $x_{B}$ by $\rho_{S}$ gives
the same entropy density. The F1 flux thus reproduces the horizon
entropy per unit boundary length in the long-interval limit \cite{Ryu:2006ef,Calabrese:2004eu}. 

\section{F1 connectivity at fixed RT entropy}

\label{sec:f1-stationary-routing}

In both the vacuum and thermal constructions of Sections \ref{sec:single-interval-F1}
and \ref{subsec:thermal}, the F1 strings that cross the interval
RT surface connect $A$ only to a restricted contact interval $B_{\mathrm{hit}}$
in the regulated complement $\bar{A}$. Their flux nevertheless gives
$S_{A}^{(0)}$, the leading von Neumann entropy of the reduced state
on $A$. In the vacuum this is the bipartite entanglement entropy
between $A$ and $\bar{A}$; in the thermal case it is not an entanglement
measure for $A$ and $\bar{A}$ alone. In neither case is it the entropy
of the pair $A$ and $B_{\mathrm{hit}}$. To understand the restricted
contact region, we vary the NS--NS fraction $\eta$. The complementary
R--R flux in the classical type IIB solution preserves the metric
\cite{Wu:2026pgc,Cagnazzo:2012se,Hoare:2013pma}. Thus the RT surface
and its capacity are fixed, while the force from $B$ on the probe
F1 changes. We use the stationary string equation and current prescription
of our previous work \cite{Wu:2026pgc} to follow this change from
the bit-thread flow at $\eta=0$ to the one-sided contacts at $\eta=1$.
The resulting on-shell currents interpolate from the two-sided geodesic
routing at $\eta=0$ to the pure NS--NS routing at $\eta=1$, where
the complementary contacts lie only in $B_{\mathrm{hit}}$. Every
member of the family gives the same leading RT entropy.

\subsection{The stationary F1 profile at mixed flux}

We first solve for the F1 string that crosses the RT surface at a
specified point. To use a static worldsheet gauge, we introduce coordinates
$\left(t_{\mathrm{R}},r,\theta\right)$ adapted to the interval Killing
boost and seek strings stationary with respect to $t_{\mathrm{R}}$.
The coordinate $t_{\mathrm{R}}$ is the interval modular (Rindler)
time, not the planar-BTZ time $t_{B}$, although their time-zero slices
coincide in the region used below. In the wedge containing $A$, $r>r_{h}$,
and the RT surface is the bifurcation curve of the interval Killing
horizon at $r=r_{h}$ \cite{Casini:2011kv}. Along the RT surface,
the coordinate $\theta\in\mathbb{R}$ increases toward the right endpoint.
We set $r_{h}=\ell_{\mathrm{AdS}}$ by a choice of coordinate normalization.
In planar BTZ, the black hole horizon is the distinct surface $z_{B}=z_{h}$.
In this section, $\eta$ denotes the background NS--NS flux fraction.
When the worldsheet integration constant of Section \ref{subsec:thermal-worldsheets}
is needed below, we denote it by $\eta_{\mathrm{ws}}$. Where the
charts overlap, the $t_{\mathrm{R}}=0$ slice agrees with $t=0$ in
the vacuum and with $t_{B}=0$ in planar BTZ. With the flux orientation
of Ref. \cite{Wu:2026pgc}, the metric and the external two-form potentials
are 
\begin{align}
dS_{S}^{2} & =-fdt_{\mathrm{R}}^{2}+\frac{dr^{2}}{f}+r^{2}d\theta^{2}, & f\left(r\right) & =\frac{r^{2}-r_{h}^{2}}{\ell_{\mathrm{AdS}}^{2}},\nonumber \\
B & =-b_{\eta}dt_{\mathrm{R}}\wedge d\theta, & b_{\eta}\left(r\right) & =\frac{\eta\left(r^{2}-r_{h}^{2}\right)}{\ell_{\mathrm{AdS}}},\nonumber \\
C^{\left(2\right)} & =-c_{\eta}dt_{\mathrm{R}}\wedge d\theta, & c_{\eta}\left(r\right) & =\frac{e^{-\phi_{0}}\sqrt{1-\eta^{2}}\left(r^{2}-r_{h}^{2}\right)}{\ell_{\mathrm{AdS}}},\qquad0\leq\eta\leq1.\label{eq:f1 stationary routing fields}
\end{align}
In particular, $H=dB=\left(2\eta/\ell_{\mathrm{AdS}}\right)rdt_{\mathrm{R}}\wedge dr\wedge d\theta$.
In the classical type IIB solution, the relative NS--NS and R--R
contributions to the curvature sum to $\eta^{2}+\left(1-\eta^{2}\right)=1$.
The two-form potentials are written in a gauge that vanishes at the
interval horizon. The bosonic F1 action (\ref{eq:f1 review action})
couples to $B$ and the probe remains at fixed internal coordinates.

As in the full horizon construction \cite{Wu:2026pgc}, take the static
worldsheet gauge $X^{t_{\mathrm{R}}}=\tau$, $X^{r}=\sigma=r$, $X^{\theta}=\theta_{\eta}\left(r\right)$
with $\epsilon^{\tau\sigma}=+1$. The induced metric is $\gamma_{\tau\tau}^{\mathrm{ind}}=-f$,
$\gamma_{\sigma\sigma}^{\mathrm{ind}}=f^{-1}+r^{2}\theta_{\eta}^{\prime2}$,
and $\gamma_{\tau\sigma}^{\mathrm{ind}}=0$, where a prime denotes
$d/dr$. Eliminating the auxiliary worldsheet metric gives the reduced
F1 action 
\begin{equation}
S_{\mathrm{F1}}=-\frac{1}{2\pi\alpha^{\prime}}\int dt_{\mathrm{R}}dr\left(\sqrt{1+fr^{2}\theta_{\eta}^{\prime2}}-b_{\eta}\theta_{\eta}^{\prime}\right).\label{eq:f1 stationary routing reduced action}
\end{equation}
The angular coordinate is cyclic, so its equation has a first integral,
with the squared equation useful for isolating the slope: 

\begin{equation}
\frac{fr^{2}\theta_{\eta}^{\prime}}{\sqrt{1+fr^{2}\theta_{\eta}^{\prime2}}}-b_{\eta}=k_{\theta},\qquad\theta_{\eta}^{\prime2}=\frac{\left(b_{\eta}+k_{\theta}\right)^{2}}{fr^{2}\left[fr^{2}-\left(b_{\eta}+k_{\theta}\right)^{2}\right]}.\label{eq:f1 stationary routing momentum}
\end{equation}
For a regular branch reaching $r_{h}$, both $b_{\eta}$ and the first
term in the first equation vanish at the horizon, so $k_{\theta}=0$.
On this branch, define the positive effective width

\begin{equation}
\mathcal{D}_{\eta}^{2}\left(r\right)\equiv r^{2}-\frac{b_{\eta}^{2}}{f}=\left(1-\eta^{2}\right)r^{2}+\eta^{2}r_{h}^{2},
\end{equation}
the radial slope and induced determinant take the form 

\begin{equation}
\theta_{\eta}^{\prime}=\frac{b_{\eta}}{fr\mathcal{D}_{\eta}}=\frac{\eta\ell_{\mathrm{AdS}}}{r\mathcal{D}_{\eta}},\qquad\sqrt{1+fr^{2}\theta_{\eta}^{\prime2}}=\frac{r}{\mathcal{D}_{\eta}}.
\end{equation}
The transverse equation and the auxiliary metric equation imply the
remaining embedding equations by worldsheet reparametrization invariance.
Using the ingoing time $t_{{\rm in}}=t_{\mathrm{R}}+\int dr/f$, the
induced metric can be written as $-fdt_{{\rm in}}^{2}+2dt_{{\rm in}}dr+r^{2}\theta_{\eta}^{\prime2}dr^{2}$
and has determinant $-1$ at $r_{h}$. The pullback of $B$ is regular
in these coordinates.

We therefore parameterize each string by its RT crossing coordinate
$\theta_{h}=\theta_{\eta}\left(r_{h}\right)$. Integrating the slope
yields 
\begin{equation}
\theta_{\eta}\left(r\right)=\theta_{h}+\frac{\ell_{\mathrm{AdS}}}{r_{h}}\log\frac{\left(1+\eta\right)r}{\mathcal{D}_{\eta}\left(r\right)+\eta r_{h}}.\label{eq:f1 stationary routing profile}
\end{equation}
At $\eta=0$ this gives $\theta_{0}\left(r\right)=\theta_{h}$, so
the time-zero strings follow geodesics normal to the cut. At $\eta=1$
the angular shift is $\left(\ell_{\mathrm{AdS}}/r_{h}\right)\log\left(r/r_{h}\right)$,
which recovers the pure NS--NS profiles. For $\eta<1$ its limiting
value is $\left(\ell_{\mathrm{AdS}}/r_{h}\right)\mathrm{arc}\tanh\eta$
as $r\to\infty$.

To follow the other end of each string, we introduce a signed coordinate
$\rho$ on the time-zero slice by $r=r_{h}\cosh\rho$, with $\rho>0$
on the side containing $A$ and $\rho<0$ on the opposite side of
the RT surface. In this notation, we define the quantity

\begin{equation}
W_{\eta}\left(\rho\right)\equiv\frac{\mathcal{D}_{\eta}\left(r_{h}\cosh\rho\right)}{r_{h}}=\sqrt{1+\left(1-\eta^{2}\right)\sinh^{2}\rho}.
\end{equation}
 On the unit AdS$_{3}$ hyperboloid $-Y_{-1}^{2}-Y_{0}^{2}+Y_{1}^{2}+Y_{2}^{2}=-1$,
the two boost wedges have 

\begin{equation}
\left(Y_{1},Y_{0}\right)=\sinh\rho\left(\cosh\frac{t_{\mathrm{R}}}{\ell_{\mathrm{AdS}}},\sinh\frac{t_{\mathrm{R}}}{\ell_{\mathrm{AdS}}}\right),\qquad\left(Y_{-1},Y_{2}\right)=\cosh\rho\left(\cosh\theta,\sinh\theta\right).
\end{equation}
With $r_{h}=\ell_{\mathrm{AdS}}$, the profile in Eq. (\ref{eq:f1 stationary routing profile})
is the intersection of the hyperboloid with the plane 
\begin{equation}
\left(1+\eta\right)e^{\theta_{h}}\left(Y_{-1}-Y_{2}\right)-\left(1-\eta\right)e^{-\theta_{h}}\left(Y_{-1}+Y_{2}\right)=\left(W_{\eta}+\eta\right)-\left(W_{\eta}-\eta\right)=2\eta.\label{eq:f1 stationary routing plane}
\end{equation}
The projected plane normal in the tangent space of AdS$_{3}$ has
squared norm $4\left(1-\eta^{2}\right)+\left(2\eta\right)^{2}=4$.
The section is therefore smooth and timelike across the boost horizons.
Analytically continuing this regular solution also continues the local
F1 equations. This fixes the two branches for each $\theta_{h}$ and
charge orientation. 

\subsection{String current and its total RT flux}

We now compute the current of the stationary solutions derived above.
At $\eta=1$, these solutions recover the pure NS-NS worldsheets of
Sections \ref{sec:single-interval-F1} and \ref{subsec:thermal}.
We keep the same cutoff and RT crossing points as we vary $\eta$,
so that we can compare the boundary contacts at fixed total RT flux.

We first derive the current before counting its boundary contacts.
The RT surface has $dA_{S}=r_{h}d\theta_{h}$ for increasing $\theta_{h}\in\left[-U,U\right]$,
where $U=U_{\varepsilon}$ in vacuum and $U=U_{\beta,\varepsilon}$
in planar BTZ. Thus $r_{h}=\ell_{\mathrm{AdS}}$ makes $\theta_{h}$
the signed RT proper distance from the midpoint in units of $\ell_{\mathrm{AdS}}$.
To match the earlier conventions, we choose $\theta_{h}=-u$ in vacuum
and $\theta_{h}=-u_{\beta}$ in planar BTZ, because the earlier coordinates
increase to the left. For $\mathcal{N}$ equally oriented sources,
set $\theta_{s}=2U/\mathcal{N}$ and place the RT crossings at $\theta_{h,k}=-U+(k+\tfrac{1}{2})\theta_{s}$,
$k=0,\ldots,\mathcal{N}-1$. The function $\theta_{\eta,k}\left(r\right)$
is given by Eq. (\ref{eq:f1 stationary routing profile}) with $\theta_{h}=\theta_{h,k}$.
Note that $\theta_{\eta,k}\left(r_{h}\right)=\theta_{h,k}$ at $r=r_{h}$
since $\mathcal{D}_{\eta}\left(r_{h}\right)=r_{h}$. The static gauge
in the normalized source current (\ref{eq:review current}) supplies
$\delta\left(t_{\mathrm{R}}-\tau\right)\delta\left(r-\sigma\right)$.
Integrating those two delta functions and using $\sqrt{-g}=r$ gives
\begin{equation}
j_{\eta}^{0r}=\frac{1}{r}\sum_{k=0}^{\mathcal{N}-1}\delta\left(\theta-\theta_{\eta,k}\left(r\right)\right),\qquad j_{\eta}^{0\theta}=\theta_{\eta}^{\prime}j_{\eta}^{0r},\qquad j_{\eta}^{r\theta}=0.\label{eq:f1 stationary routing discrete current}
\end{equation}
At fixed $r$ the label $\theta_{h}=\theta-\left[\theta_{\eta}\left(r\right)-\theta_{\eta}\left(r_{h}\right)\right]$
differs from $\theta$ by a common shift, so the delta functions carry
a unit Jacobian. In the large $\mathcal{N}$ limit, the weighted sum
of delta functions $\theta_{s}\sum_{k}\delta\left(\theta_{h}-\theta_{h,k}\right)$
gives a uniform density of one on $\left[-U,U\right]$ and zero outside,
as seen by integrating it against any smooth function.

The slice metric is $h_{ij}dx^{i}dx^{j}=dr^{2}/f+r^{2}d\theta^{2}$
and $N_{\mathrm{ADM}}=\sqrt{f}$. The projected components and their
norm inside the bundle are 

\begin{equation}
q_{\eta}^{r}=N_{\mathrm{ADM}}j_{\eta}^{0r}=\frac{\mathcal{N}\sqrt{f}}{2Ur},\qquad q_{\eta}^{\theta}=N_{\mathrm{ADM}}j_{\eta}^{0\theta}=\theta_{\eta}^{\prime}q_{\eta}^{r}.\label{eq:f1 stationary routing charge}
\end{equation}
Their norm is

\begin{equation}
|q_{\eta}|^{2}=\left(\frac{\mathcal{N}}{2Ur}\right)^{2}\left(1+fr^{2}\theta_{\eta}^{\prime2}\right)=\left(\frac{\mathcal{N}}{2U\mathcal{D}_{\eta}}\right)^{2}.
\end{equation}
Since $\sqrt{h}=r/\sqrt{f}$, $\sqrt{h}q_{\eta}^{r}$ is constant
and $\sqrt{h}q_{\eta}^{\theta}$ is independent of $\theta$ in the
bundle. Its sides are profiles of constant $\theta_{h}$, so there
is no normal flux across the sides and the extension by zero preserves
$\nabla_{i}q_{\eta}^{i}=0$. On a segment of fixed $r$ spanning the
populated labels, the outward unit normal is $\sqrt{f}\partial_{r}$
and the line element is $rd\theta$. Its charge density flux is given
by
\begin{equation}
\Phi\left[q_{\eta}\right]=\int_{-U}^{U}\frac{q_{\eta}^{r}}{\sqrt{f}}rd\theta_{h}=\frac{\mathcal{N}}{2U}\int_{-U}^{U}d\theta_{h}=\mathcal{N}.\label{eq:f1 stationary routing charge flux}
\end{equation}
Here $d\theta=d\theta_{h}$ at fixed $r$, and the formula also gives
the flux at the RT cut by continuity.

For $\eta<1$ the charge norm varies along each string. A pointwise
rescaling to norm $\rho_{S}$ would give 
\begin{equation}
\nabla_{i}\left(\rho_{S}\frac{q_{\eta}^{i}}{|q_{\eta}|}\right)=\frac{\rho_{S}\sqrt{f}}{r}\mathcal{D}_{\eta}^{\prime}=\frac{\rho_{S}\left(1-\eta^{2}\right)\sqrt{f}}{\mathcal{D}_{\eta}}\label{eq:f1 stationary routing unit divergence}
\end{equation}
in the bundle of the $A$ wedge. Instead, as in Ref. \cite{Wu:2026pgc},
we use a single constant $C_{\mathrm{geom}}$ and fix it by the local
capacity: 
\begin{equation}
C_{\mathrm{geom}}=\inf_{r\geq r_{h}}\frac{\rho_{S}}{|q_{\eta}|}=\frac{2Ur_{h}\rho_{S}}{\mathcal{N}}.\label{eq:f1 stationary routing conversion}
\end{equation}
This uses $\mathcal{D}_{\eta}^{2}-r_{h}^{2}=\left(1-\eta^{2}\right)\left(r^{2}-r_{h}^{2}\right)\geq0$
and the presence of the cut $r=r_{h}$ in the support. The resulting
entropy flow in the $A$ wedge is 

\begin{equation}
v_{\mathrm{phys},\eta}^{i}=C_{\mathrm{geom}}q_{\eta}^{i}=\rho_{S}\frac{r_{h}\sqrt{f}}{r}\left(1,\frac{\eta\ell_{\mathrm{AdS}}}{r\mathcal{D}_{\eta}}\right).
\end{equation}
It is easy to check that

\begin{equation}
\nabla_{i}v_{\mathrm{phys},\eta}^{i}=0,\qquad|v_{\mathrm{phys},\eta}|=\rho_{S}\frac{r_{h}}{\mathcal{D}_{\eta}}\leq\rho_{S}.
\end{equation}
The total RT flux is therefore 

\begin{equation}
S_{A}^{\left(0\right)}=C_{\mathrm{geom}}\mathcal{N}=2\rho_{S}r_{h}U=2\rho_{S}\ell_{\mathrm{AdS}}U.
\end{equation}
The total is independent of $\eta$. 

\subsection{Vacuum: from two-sided to one-sided contact}

Now, we fix the cutoff and the RT crossing points while varying $\eta$.
Each profile has one cutoff contact $x_{A}$ in $A_{\varepsilon}$.
We ask whether its other contact $x_{\mathrm{out}}$ lies to the left
or right of $A_{\varepsilon}$.

For $0\leq\eta<1$, write $\vartheta=\mathrm{arc}\tanh\eta$ and $w=\theta_{h}+\vartheta$.
At $t=0$, the Poincare coordinates give $Y_{-1}=\left(R^{2}+x^{2}+z^{2}\right)/\left(2Rz\right)$
and $Y_{2}=x/z$. Eq. (\ref{eq:f1 stationary routing plane}) becomes
\begin{equation}
\sinh w\frac{R^{2}+x^{2}+z^{2}}{2Rz}-\cosh w\frac{x}{z}=\sinh\vartheta.\label{eq:f1 stationary routing vacuum plane}
\end{equation}
At $z=\varepsilon$, this is a quadratic in $x$ for $w\ne0$. The
sum of its roots is $2R\coth w$. Since $|x_{A}|<R$ and $|\coth w|>1$,
we have 
\begin{equation}
x_{\mathrm{out}}=2R\coth w-x_{A}\qquad\Longrightarrow\qquad\begin{cases}
x_{\mathrm{out}}<-R, & w<0,\\
x_{\mathrm{out}}>R, & w>0.
\end{cases}
\end{equation}
For a string with fixed $\theta_{h}$, increasing $\eta$ increases
$w$. When a string switches sides, $x_{\mathrm{out}}$ tends to $-\infty$
as $w\to0^{-}$ and to $+\infty$ as $w\to0^{+}$. The profile with
$w=0$ reaches Poincare infinity. A single crossing point contributes
no flux in the continuum limit.

The dividing point on the RT surface is at $\theta_{h}=-\vartheta$.
Strings crossing at smaller $\theta_{h}$ connect to the left of $A_{\varepsilon}$,
whereas those crossing at larger $\theta_{h}$ connect to the right.
As $\eta$ increases, the dividing point moves left from the midpoint
of the fixed range $\left[-U_{\varepsilon},U_{\varepsilon}\right]$.
The left flux fraction is therefore the length of the part of this
range below $-\vartheta$, divided by the total length: 
\begin{equation}
\frac{\Phi_{\mathrm{L}}}{S_{A}^{\left(0\right)}}=\frac{\left[U_{\varepsilon}-\vartheta\right]_{+}}{2U_{\varepsilon}},\qquad U_{\varepsilon}=\mathrm{arc}\cosh\frac{R}{\varepsilon},\qquad\left[y\right]_{+}=\max\left(y,0\right).\label{eq:f1 stationary routing vacuum split}
\end{equation}
At $\eta=0$, half the flux goes to each side. The left flux vanishes
when the dividing point reaches $\theta_{h}=-U_{\varepsilon}$, at
\begin{equation}
\eta_{\mathrm{crit}}=\tanh U_{\varepsilon}=\sqrt{1-\varepsilon^{2}/R^{2}}.\label{eq:f1 stationary routing threshold}
\end{equation}
At this value the profile through the left endpoint of the regulated
RT curve still reaches infinity. For $\eta>\eta_{\mathrm{crit}}$,
all complementary contacts lie in a bounded interval on the right.
In the pure NS--NS case $\eta=1$, the profiles obey Eq. (\ref{eq:circle})
with $z_{0}=Re^{-\theta_{h}}$. Their two cutoff contacts are related
by reflection about $x=R$: 
\begin{equation}
x_{\mathrm{out}}=2R-x_{A}.\label{eq:f1-stationary-routing-vacuum-reflection}
\end{equation}
The interval of complementary contacts is then $B_{\mathrm{hit},\varepsilon}$.
Eq. (\ref{eq:f1 stationary routing vacuum split}) also shows that
the limits do not commute: at any fixed $\eta<1$, removing the cutoff
sends $U_{\varepsilon}\to\infty$ and restores one half of the flux
on each side, whereas taking $\eta\to1$ first gives the one-sided
pure NS--NS routing.

\begin{figure}[h]
\includegraphics[width=1\linewidth]{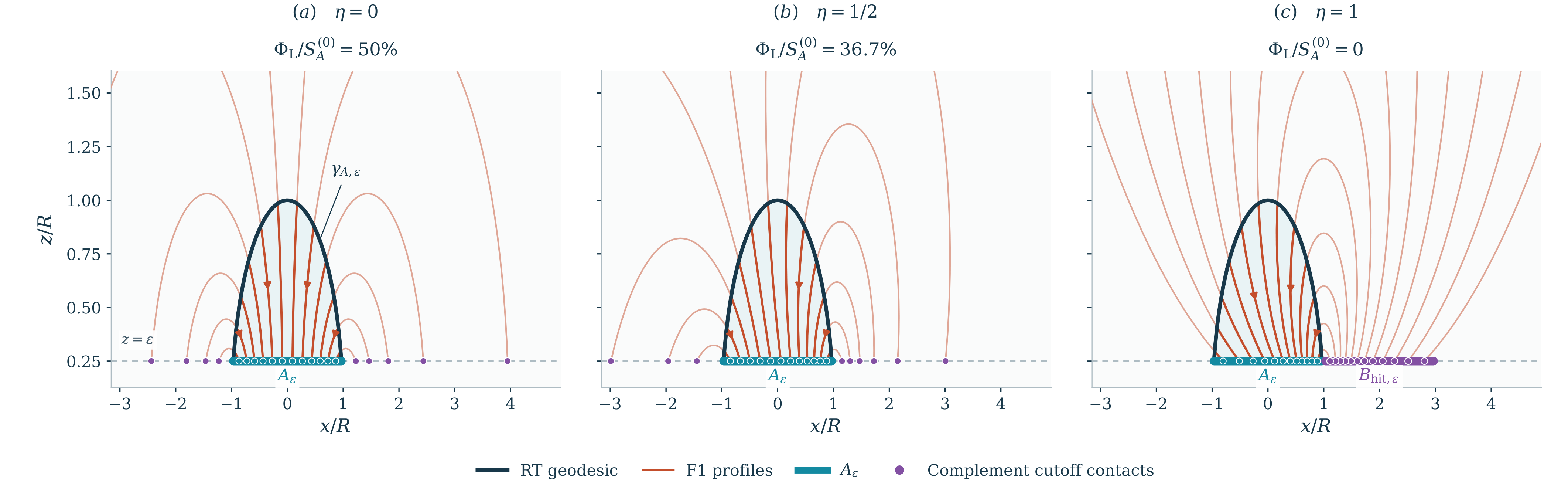}
\caption{Vacuum F1 profiles for $\eta=0,\,\tfrac{1}{2},\,1$ at $t=0$ and
$\varepsilon/R=0.25$. The same RT surface $\gamma_{A,\varepsilon}$
(navy) and crossing points $\theta_{h,k}$ are used in all panels.
Violet dots mark the complementary cutoff contacts, and cyan marks
$A_{\varepsilon}$. The contacts lie on both sides at $\eta=0$ and
$\eta=1/2$, and only on the right at $\eta=1$. Some sampled arcs
leave the plotted window.}
\label{fig:f1-vacuum-flux-triptych}
\end{figure}

For the cutoff used in Fig. (\ref{fig:f1-vacuum-flux-triptych}),
$\eta_{\mathrm{crit}}\simeq0.968$. The middle panel at $\eta=1/2$
therefore still has contacts on both sides, but more flux goes to
the right. The total RT flux is the same in all three panels.

\subsection{Planar BTZ: from inter-exterior flux to one-sided contact}

As in vacuum, we follow each string from $A$ to its other end. In
BTZ, that end can lie to the left or right of $A$ in the original
exterior, or it can reach the second exterior of the extended geometry
\cite{Maldacena:2001kr}.

Let us keep $d=L_{A}/\left(2z_{h}\right)$, $\xi=\left(x_{B}-x_{c}\right)/z_{h}$,
and $W=z_{h}/z_{B}$. On the extended time-zero slice, $\mathcal{Q}=\pm\sqrt{W^{2}-1}$
is positive in the exterior which contains $A$, negative in the second
exterior, and zero on the BTZ bifurcation line. The coordinate transformation
is given by

\begin{equation}
\mathcal{Q}=\frac{Y_{-1}+\cosh dY_{1}}{\sinh d},\qquad Y_{2}=W\sinh\xi,\qquad Y_{-1}=\frac{W\cosh d\cosh\xi-\mathcal{Q}}{\sinh d}.\label{eq:f1 stationary routing btz map}
\end{equation}
 In the original exterior, $\mathcal{Q}=Wh_{B}$, so $Y_{1}=0$ gives
the thermal RT curve $\cosh\xi=h_{B}\cosh d$.

For $\eta<1$, the angular coordinate in Eq. (\ref{eq:f1 stationary routing profile})
approaches $w=\theta_{h}+\vartheta$ at the other end, where $\vartheta=\mathrm{arc}\tanh\eta$
as in vacuum. Taking $\rho\to-\infty$ in the first relation of Eq.
(\ref{eq:f1 stationary routing btz map}) gives 
\begin{equation}
\lim_{\rho\to-\infty}\frac{\mathcal{Q}}{\cosh\rho}=\frac{\cosh w-\cosh d}{\sinh d}.\label{eq:f1 stationary routing btz sign}
\end{equation}
This limit is negative for $|w|<d$, so the string reaches the second
exterior. For $|w|>d$, it ends in the original exterior, where $Y_{2}=W\sinh\xi$
gives $\sinh\xi\to\sinh d\sinh w/\left(\cosh w-\cosh d\right)$. Therefore
$|\xi|>d$ there, and the sign of $w$ determines the side of $A$.
The other end lies 
\begin{equation}
\begin{cases}
\text{to the left of }A\text{ in the original exterior}, & w<-d,\\
\text{in the second exterior}, & -d<w<d,\\
\text{to the right of }A\text{ in the original exterior}, & w>d.
\end{cases}
\end{equation}
The two values $w=\pm d$ correspond to two individual string profiles
and therefore do not affect the continuum flux.

\subsubsection{Net horizon flux and extensive thermal entropy}

The endpoint criterion tells us which strings connect the two exteriors.
To determine how much of the total flow into $A$ is carried by strings
connecting the two exteriors, we evaluate the same F1 current across
the BTZ bifurcation line on the $t_{B}=0$ slice.

The definition of the string current supplies the required signed
counting rule \cite{Zwiebach:2004tj}. Eq. (\ref{eq:f1 stationary routing charge flux})
gives $\Phi\left[q_{\eta}\right]=\mathcal{N}$ on the regulated RT
curve, where all $\mathcal{N}$ strings cross once with the same orientation.
On a general oriented curve the count is signed: two crossings with
opposite orientations contribute $+1$ and $-1$ in string charge
units. The same rule applies to two crossings of a single string.
Since $v_{\mathrm{phys},\eta}^{\,i}=C_{\mathrm{geom}}q_{\eta}^{i}$
with constant $C_{\mathrm{geom}}$, the entropy flow inherits this
counting rule. We therefore define 

\begin{equation}
\Phi_{\mathrm{hor}}^{\mathrm{net}}\equiv\int_{\mathcal{H}_{B}}v_{\mathrm{phys},\eta}^{i}n_{i}ds,\qquad\mathcal{H}_{B}=\{\mathcal{Q}=0\},
\end{equation}
 where $n_{i}$ is the unit normal pointing from the second exterior
toward the one containing $A$, and $ds$ is the proper length element
along the line.

\begin{figure}[h]
\includegraphics[width=1\linewidth]{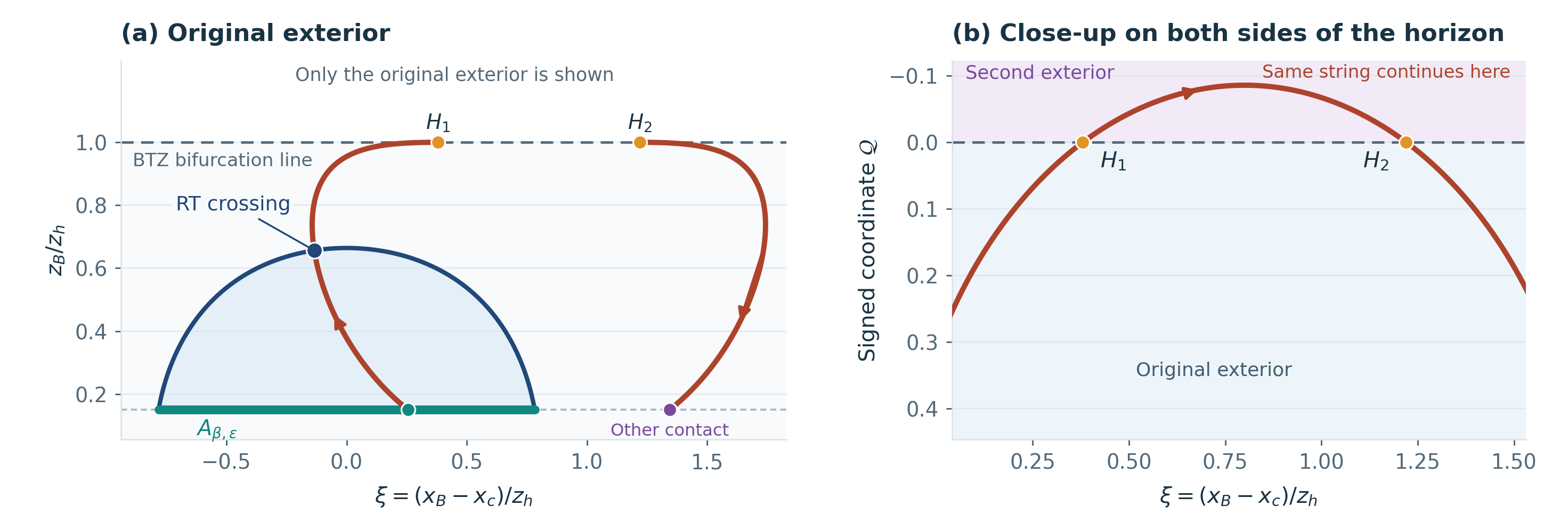} \caption{A returning F1 string at $\eta=1$, with $d=0.80$, $\varepsilon/z_{h}=0.15$,
and $\theta_{h}=-U_{\beta,\varepsilon}/12$. The left panel shows
its single RT crossing and the two intersections $H_{1},H_{2}$ with
the BTZ bifurcation line. The right panel uses the signed coordinate
$\mathcal{Q}$ to show the segment joining $H_{1}$ and $H_{2}$ in
the second exterior. The original exterior is below the line in both
panels. The arrows trace the profile from $A$. The two horizon crossings
carry equal and opposite flux.}
\label{fig:f1-btz-rt-vs-horizon}
\end{figure}

A string that joins the two exteriors has one net crossing toward
$A$. If both ends are in the original exterior, any passage into
the second exterior is paired with a passage back. The same conserved
charge crosses in opposite directions, so the two contributions to
$\Phi_{\mathrm{hor}}^{\mathrm{net}}$ cancel. Fig. (\ref{fig:f1-btz-rt-vs-horizon})
illustrates the cancellation for a single on-shell F1 string. It crosses
the BTZ bifurcation line twice and the interval RT curve only once.

The strings in the ensemble cross the regulated RT curve at $\theta_{h}\in\left[-U,U\right]$,
with $U=U_{\beta,\varepsilon}$. We follow their full profiles without
imposing a separate cutoff in the second exterior. For $\eta<1$ the
endpoint criterion $|w|<d$ is equivalent to $-d-\vartheta<\theta_{h}<d-\vartheta$.
Each such string contributes its RT flux to the net horizon flux,
and all other strings give zero net contribution. The net flux is
therefore 
\begin{equation}
\Phi_{\mathrm{hor}}^{\mathrm{net}}=\rho_{S}\ell_{\mathrm{AdS}}\mathrm{Length}\left(\left[-U,U\right]\cap\left[-d-\vartheta,d-\vartheta\right]\right),\qquad U=U_{\beta,\varepsilon}.\label{eq:f1 stationary routing btz split}
\end{equation}
For $U>d$, the overlap length is explicitly 
\begin{equation}
\mathrm{Length}=\begin{cases}
2d, & 0\leq\vartheta\leq U-d,\\
U+d-\vartheta, & U-d<\vartheta<U+d,\\
0, & \vartheta\geq U+d.
\end{cases}\label{eq:f1 stationary routing btz overlap piecewise}
\end{equation}
If $U\ge d+\vartheta$, the entire range of crossings leading to the
second exterior lies within $\left[-U,U\right]$, so the overlap has
length $2d$. Using $2d=L_{A}/z_{h}$, $\beta=2\pi z_{h}$, and $c=6\rho_{S}\ell_{\mathrm{AdS}}$,
we obtain 
\begin{equation}
\Phi_{\mathrm{hor}}^{\mathrm{net}}=2d\rho_{S}\ell_{\mathrm{AdS}}=s_{\mathrm{th}}L_{A}=\frac{\pi cL_{A}}{3\beta},\qquad s_{\mathrm{th}}=\frac{\pi c}{3\beta}.\label{eq:f1 stationary routing btz horizon}
\end{equation}
The horizon has proper length $\ell_{\mathrm{AdS}}/z_{h}$ per unit
boundary length, giving the thermal entropy density $s_{\mathrm{th}}=\rho_{S}\ell_{\mathrm{AdS}}/z_{h}$.
The net flux in this regime therefore reproduces the extensive thermal
entropy of a segment of length $L_{A}$. At $\eta=0$ it agrees with
the bit-thread construction \cite{Mintchev:2022fcp,Caggioli:2024uza}.

The standard bit-thread flow is also signed, so opposite passages
through an auxiliary curve cancel for a conserved flow \cite{Freedman:2016zud}.
A general bit-thread flow can have a returning profile. The geodesic
construction in planar BTZ excludes such a return because the bifurcation
line is itself a geodesic and two distinct geodesics on this slice
meet at most once. The string charge density gives a physical origin
to the signed count and ties each counted profile to the F1 equation.
At fixed metric and dilaton the bit-thread constraints have no $\eta$
dependence. The Kalb--Ramond coupling in the probe equation, by contrast,
predicts how the string connections change with $\eta$.

As $\eta$ rises, $w=\theta_{h}+\vartheta$ increases at each fixed
$\theta_{h}$. A string whose complementary end is to the left can
come to reach the second exterior and then end to the right. At fixed
cutoff and $U>d$, the range $\left[-d-\vartheta,d-\vartheta\right]$
moves to smaller $\theta_{h}$. Its full length contributes until
its left endpoint reaches $-U$ at $\eta=\tanh\left(U-d\right)$.
Although the identities of the strings reaching the second exterior
have changed, their total signed flux remains equal to Eq. (\ref{eq:f1 stationary routing btz horizon}).
Beyond this point, the overlap has length $U+d-\vartheta$ until its
right endpoint also reaches $-U$, at $\eta=\tanh\left(U+d\right)$.
The net flux thus vanishes. Above this threshold, for $\eta<1$, every
complementary end is to the right of $A$ in the original exterior.

At the pure NS--NS point $\eta=1$, the profile is $\theta=\theta_{h}+\log\cosh\rho$.
The growth of $Y_{-1}$ as $\cosh^{2}\rho$ at large $|\rho|$ makes
$\mathcal{Q}$ positive at both ends. In the original exterior, Eq.
(\ref{eq:f1 stationary routing plane}) becomes $W\left[\cosh\left(\xi-d\right)-h_{B}\right]=\sinh de^{-\theta_{h}}$.
This is the pure NS--NS surface (\ref{eq:BTZ surface}) at $t_{B}=0$
with $\eta_{\mathrm{ws}}=\sinh de^{-\theta_{h}}$, where $\eta_{\mathrm{ws}}$
is the integration constant called $\eta$ in Section \ref{subsec:thermal-worldsheets}.
The two cutoff contacts have the same $W$ and $h_{B}$. Since $\cosh\left(\xi-d\right)$
is even about $\xi=d$, the contacts are reflected about $x_{2}$,
and both ends are in the same exterior: 
\begin{equation}
x_{B,\mathrm{out}}=2x_{2}-x_{B,A},\qquad\Phi_{\mathrm{hor}}^{\mathrm{net}}=0.\label{eq:f1 stationary routing btz reflection}
\end{equation}
All complementary cutoff contacts are then in $B_{\mathrm{hit},\varepsilon}^{\beta}$.

\begin{figure}[h]
\includegraphics[width=1\linewidth]{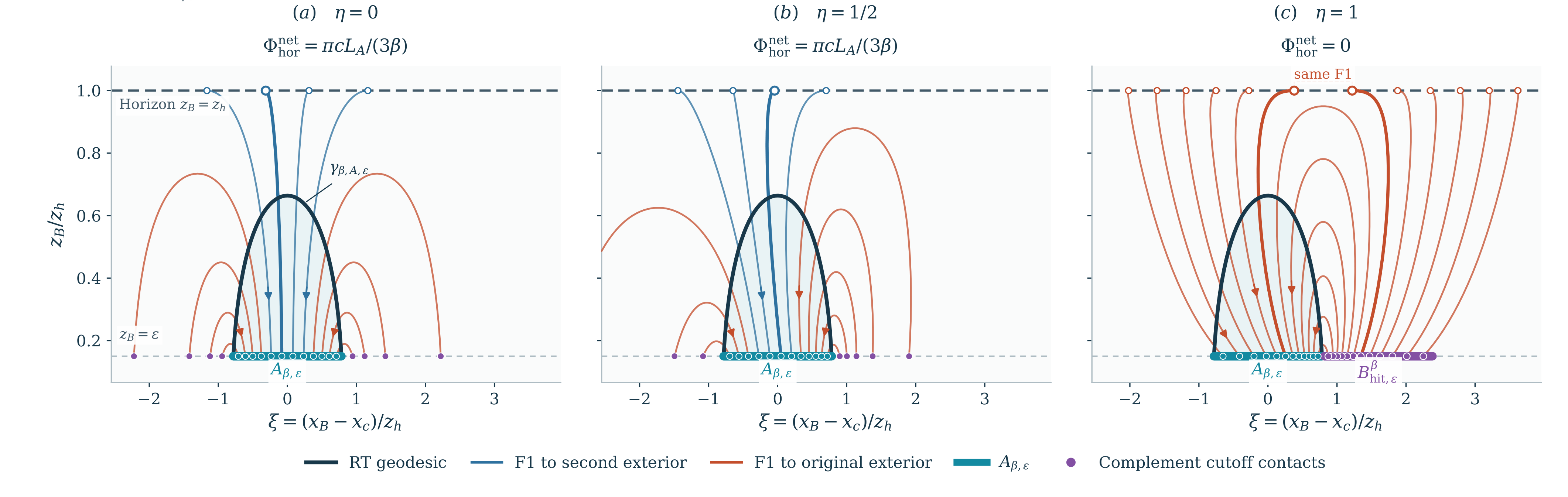}
\caption{Thermal F1 sources in the original planar BTZ exterior at $t_{B}=0$
for $\eta=0,\,\tfrac{1}{2},\,1$. All panels use $d=0.80$, $\varepsilon/z_{h}=0.15$,
and the same crossing points $\theta_{h,k}$ on $\gamma_{\beta,A,\varepsilon}$.
Blue strings reach the boundary of the second exterior. Red strings
have both ends in the original exterior. Panels (a) and (b) have net
horizon flux $s_{\mathrm{th}}L_{A}$, although different crossing
points connect the two exteriors. In panel (c), every complementary
cutoff contact lies in $B_{\mathrm{hit},\varepsilon}^{\beta}$ and
the net horizon flux is zero. The two highlighted branches in (c)
belong to one string that crosses the bifurcation line and returns.}
\label{fig:f1-btz-flux-triptych}
\end{figure}

The pure NS--NS profile also specifies which strings pass into the
second exterior and return. Setting $s=-\sinh\rho>0$ on the $\rho<0$
branch, Eq. (\ref{eq:f1 stationary routing btz map}) gives 
\begin{equation}
2\sinh d\mathcal{Q}=e^{\theta_{h}}s^{2}-2s\cosh d+e^{\theta_{h}}+e^{-\theta_{h}}.
\end{equation}
The discriminant is $4\left(\sinh^{2}d-e^{2\theta_{h}}\right)$, so
there are two positive roots if $\theta_{h}<\log\sinh d$. Between
them the string is in the second exterior, with one crossing in each
direction. Both roots lie beyond its single RT crossing at $\rho=0$.
Fig. (\ref{fig:f1-btz-flux-triptych}) compares how the boundary connections
change as $\eta$ increases, with the same regulated RT curve throughout.

Removing the cutoff at fixed $\eta<1$ sends $U$ to infinity and
recovers Eq. (\ref{eq:f1 stationary routing btz horizon}). Taking
$\eta\to1$ first at fixed cutoff gives zero net horizon flux instead.
The total RT flux still gives the same interval entropy (\ref{eq:BTZ CFT entropy})
for every $\eta$. The $\eta$ dependence is a property of the physical
F1 current: it specifies which parts of the complement receive the
fixed RT flux. In particular, the pure NS--NS contacts lie in $B_{\mathrm{hit},\varepsilon}^{\beta}$,
while the partial trace defining the entropy of $A$ includes its
full complement. 

\section{Conclusion and discussion}

\label{sec:physical-interpretation}

In this paper, we constructed families of on-shell F1 probes for a
single interval in the AdS$_{3}$ vacuum and in planar BTZ. On the
time-zero slice, the entropy flow obtained from their conserved spatial
Kalb--Ramond current saturates the string-frame capacity $\rho_{S}$
in the normal direction along the RT surface. Its total flux therefore
reproduces the standard leading vacuum and thermal interval entropies.
The worldsheet embeddings also specify where the strings carrying
this flux meet the cutoff boundary.

The mixed-flux calculation then separates the total RT flux from the
string connections. At fixed metric and dilaton, the geometric bit-thread
constraints are independent of the NS--NS fraction $\eta$. The Kalb--Ramond
coupling in the F1 equation changes with $\eta$ and, in the boost-invariant
sector, selects a regular stationary profile for each RT crossing.
At $\eta=0$ these profiles give the geodesic bit-thread flow. In
vacuum, the flux to the left of $A$ vanishes at the finite-cutoff
threshold $\eta_{\mathrm{crit}}=\tanh U_{\varepsilon}$.

In planar BTZ, the net flux through the bifurcation line equals $s_{\mathrm{th}}L_{A}$
when $\eta<1$ and $U_{\beta,\varepsilon}\geq d+\vartheta$, where
$\vartheta=\mathrm{arc}\tanh\eta$. This condition is eventually satisfied
when the cutoff is removed at any fixed $\eta<1$. At fixed cutoff,
the overlap of RT crossings that connect the two exteriors vanishes
at $\eta=\tanh\left(U_{\beta,\varepsilon}+d\right)$. At the pure
NS--NS point, $\eta=1$, both cutoff contacts lie in the original
exterior and the net horizon flux is zero, while the thermal entropy
density and total RT flux remain unchanged. A spatial profile can
still enter the second exterior and return: the two horizon crossings
contribute with opposite signs, while the string crosses the interval
RT surface only once.

The total entropy flux is thus fixed by the RT surface, while the
F1 dynamics determine how that flux is distributed among complementary
regions. We now distinguish the density of cutoff contacts from their
support, and explain what the support means for the boundary subsystem.

\subsubsection*{String ensemble and attachment density}

\label{subsec:f1-population-independence}

We now clarify whether the normalized F1 flow depends on how the strings
are distributed along the RT surface. Keeping the string profiles
and the range of crossing points fixed, we examine how reweighting
the ensemble affects two quantities: the entropy flux through the
interval and the density of string contacts at the cutoff.

Let $\chi_{0}$ denote the RT crossing label, with $\chi_{0}=u$ in
vacuum and $\chi_{0}=u_{\beta}$ in planar BTZ. The RT length element
is $dA_{S}=\ell_{\mathrm{AdS}}|d\chi_{0}|$. Replacing the uniform
string density with a smooth positive density $\nu\left(\chi_{0}\right)$
on the same label range multiplies the projected current by $\nu\left(\chi_{0}\right)/\nu_{\mathrm{uniform}}$.
This factor is constant along each string profile, so the reweighted
current is still conserved. In pure NS--NS flux, the factor cancels
in $\widehat{q}^{i}=q^{i}/|q|$, leaving $v_{\mathrm{phys}}^{i}=\rho_{S}\widehat{q}^{i}$
and its maximal flux unchanged.

Reweighting the ensemble changes the density of cutoff contacts on
the regulated interval. Writing $x_{\partial}=x$ in vacuum and $x_{\partial}=x_{B}$
in planar BTZ, the number of contacts per unit cutoff coordinate is
\begin{equation}
\frac{d\mathcal{N}_{A}}{dx_{\partial}}=\nu\left(\chi_{0}\left(x_{\partial}\right)\right)\left|\frac{d\chi_{0}}{dx_{\partial}}\right|.\label{eq:f1 boundary density jacobian}
\end{equation}
Even a uniform density of strings along the RT surface gives a nonuniform
contact density when this Jacobian varies. In vacuum, the normalized
flow carries $\rho_{S}\ell_{\mathrm{AdS}}|du|$ per label interval.
Evaluating Eq. (\ref{eq:inverse}) at $t=0$ and $z=\varepsilon$
therefore gives 
\begin{equation}
\frac{d\Phi_{A}}{dx_{A}}=\rho_{S}\ell_{\mathrm{AdS}}\frac{2\left(R-x_{A}\right)}{\left(R-x_{A}\right)^{2}+\varepsilon^{2}},\qquad-a_{\varepsilon}\leq x_{A}\leq a_{\varepsilon}.\label{eq:f1-info-f1-boundary-flux}
\end{equation}
This is the entropy flux per unit cutoff coordinate of the chosen
normalized F1 maximizer. Positive reweighting changes the unnormalized
string-contact density but cancels from $q^{i}/|q|$, so this normalized
entropy-flux density remains unchanged.

\subsubsection*{Cutoff contacts and the reduced state}

\label{subsec:f1-boundary-support}

We now turn from how densely the strings meet the cutoff to which
parts of the complement they reach. Does the region reached by the
chosen F1 current coincide with the region traced out to obtain $\rho_{A}$?
To clarify what the contact region tells us about boundary correlations,
we examine whether $A$ can have nonzero mutual information with a
region that receives no flux from this current.

In pure NS--NS flux, the reflection maps (\ref{eq:f1-stationary-routing-vacuum-reflection})
and~(\ref{eq:f1 stationary routing btz reflection}) give the complementary
cutoff contacts in vacuum and planar BTZ, respectively: 

\begin{equation}
B_{\mathrm{hit},\varepsilon}=\left[2R-a_{\varepsilon},2R+a_{\varepsilon}\right],\qquad B_{\mathrm{hit},\varepsilon}^{\beta}=\left[2x_{2}-b_{\beta,\varepsilon},2x_{2}-a_{\beta,\varepsilon}\right].\label{eq:f1 info contact intervals}
\end{equation}
These intervals specify where the chosen F1 current reaches the regulated
complement. The reduced state $\rho_{A}=\mathrm{Tr}_{\bar{A}}\rho$
is defined using the full complement $\bar{A}$ on the boundary line.
In the vacuum, that complement purifies $A$. In the thermal case,
the second boundary in the two-sided description purifies the thermal
state on the original boundary line \cite{Maldacena:2001kr}. Starting
from that purified state, one traces over both the second boundary
and $\bar{A}$ to obtain $\rho_{A}$. The contact intervals are supports
of this current, not physical boundaries of the CFT. They therefore
impose no conformal boundary condition and introduce no boundary entropy
$\log g$ \cite{Calabrese:2004eu,Affleck:1991tk,Takayanagi:2011zk}.

The vacuum provides a direct example of the distinction. The interval
$C_{0}=\left(-7R/2,-3R/2\right)$ lies outside $B_{\mathrm{hit},\varepsilon}$
and receives no flux from this F1 maximizer. For $A\cup C_{0}$, the
disconnected and connected RT pairings give products of endpoint separations
$4R^{2}$ and $9R^{2}/4$, respectively. The connected pairing has
the smaller length in the continuum limit \cite{Headrick:2010zt},
so 
\begin{equation}
I^{\left(0\right)}\left(A:C_{0}\right)=\frac{c}{3}\log\frac{16}{9}>0.\label{eq:f1 info vacuum mutual information}
\end{equation}
Thus $A$ can share leading boundary correlations with a region that
receives no flux from the chosen F1 current. The cutoff contacts describe
how this current realizes $S_{A}^{\left(0\right)}$, and the full
complement determines the trace that defines $\rho_{A}$. 

\vspace{5mm}

\noindent {\bf Acknowledgements} 
HW is supported by NSFC Grant No.12105191. SY is supported by NSFC Grant No.12105031 and No.12547101.


\begin{thebibliography}{99}

%\cite{Ryu:2006bv} 
\bibitem{Ryu:2006bv} S.~Ryu and T.~Takayanagi,   ``Holographic derivation of entanglement entropy from AdS/CFT,''   Phys.\ Rev.\ Lett.\  {\bf 96}, 181602 (2006)   doi:10.1103/PhysRevLett.96.181602   [hep-th/0603001].   %%CITATION = doi:10.1103/PhysRevLett.96.181602;%%   %1197 citations counted in INSPIRE as of 22 Mar 2017

%\cite{Ryu:2006ef} 
\bibitem{Ryu:2006ef}    S.~Ryu and T.~Takayanagi,   ``Aspects of Holographic Entanglement Entropy,''   JHEP {\bf 0608}, 045 (2006)   doi:10.1088/1126-6708/2006/08/045   [hep-th/0605073].   %%CITATION = doi:10.1088/1126-6708/2006/08/045;%%   %846 citations counted in INSPIRE as of 16 Oct 2017


%bit threads%

%\cite{Freedman:2016zud} 
\bibitem{Freedman:2016zud} M.~Freedman and M.~Headrick, ``Bit threads and holographic entanglement,'' Commun. Math. Phys. \textbf{352}, no.1, 407-438 (2017) doi:10.1007/s00220-016-2796-3 [arXiv:1604.00354 [hep-th]]. %206 citations counted in INSPIRE as of 16 Aug 2025

%\cite{Headrick:2017ucz} 
\bibitem{Headrick:2017ucz} M.~Headrick and V.~E.~Hubeny, ``Riemannian and Lorentzian flow-cut theorems,'' Class. Quant. Grav. \textbf{35}, no.10, 105012 (2018) doi:10.1088/1361-6382/aab83c [arXiv:1710.09516 [hep-th]]. %86 citations counted in INSPIRE as of 04 Aug 2026


%\cite{Agon:2018lwq} 
\bibitem{Agon:2018lwq} C.~A.~Ag{\'o}n, J.~De Boer and J.~F.~Pedraza, ``Geometric Aspects of Holographic Bit Threads,'' JHEP \textbf{05}, 075 (2019) doi:10.1007/JHEP05(2019)075 [arXiv:1811.08879 [hep-th]]. %89 citations counted in INSPIRE as of 24 Aug 2026




%\cite{Hubeny:2018bri} 
\bibitem{Hubeny:2018bri} V.~E.~Hubeny, ``Bulk locality and cooperative flows,'' JHEP \textbf{12}, 068 (2018) doi:10.1007/JHEP12(2018)068 [arXiv:1808.05313 [hep-th]]. %44 citations counted in INSPIRE as of 17 Nov 2026


 %\cite{Mintchev:2022fcp} 
\bibitem{Mintchev:2022fcp} M.~Mintchev and E.~Tonni, ``Modular conjugations in 2D conformal field theory and holographic bit threads,'' JHEP \textbf{12}, 149 (2022) doi:10.1007/JHEP12(2022)149 [arXiv:2209.03242 [hep-th]]. %20 citations counted in INSPIRE as of 17 Nov 2026


%\cite{Caggioli:2024uza} 
\bibitem{Caggioli:2024uza} S.~Caggioli, F.~Gentile, D.~Seminara and E.~Tonni, ``Holographic thermal entropy from geodesic bit threads,'' JHEP \textbf{07}, 088 (2024) doi:10.1007/JHEP07(2024)088 [arXiv:2403.03930 [hep-th]]. %10 citations counted in INSPIRE as of 24 Aug 2026


%%our work%%

%\cite{Wu:2025qwc} 
\bibitem{Wu:2025qwc} H.~Wu and S.~Ying, ``Toward a worldsheet theory of entanglement entropy,'' Phys. Rev. D \textbf{113}, no.6, 066006 (2026) doi:10.1103/cgg9-p6xy [arXiv:2511.16586 [hep-th]]. %2 citations counted in INSPIRE as of 04 Aug 2026



%\cite{Wu:2026qha} 
\bibitem{Wu:2026qha} H.~Wu and S.~Ying, ``String charge density/bit threads correspondence and its S-duality in type IIB supergravity,'' [arXiv:2609.15929 [hep-th]]. %1 citations counted in INSPIRE as of 13 Oct 2026


%\cite{Wu:2026kci} 
\bibitem{Wu:2026kci} H.~Wu and S.~Ying, ``T-duality of string charge density/bit threads correspondence in type II supergravity,'' [arXiv:2609.24366 [hep-th]]. %0 citations counted in INSPIRE as of 13 Oct 2026



%\cite{Tseytlin:1996qg} 
\bibitem{Tseytlin:1996qg} A.~A.~Tseytlin, ``Extremal black hole entropy from conformal string sigma model,'' Nucl. Phys. B \textbf{477}, 431-448 (1996) doi:10.1016/0550-3213(96)00383-5 [arXiv:hep-th/9605091 [hep-th]]. %49 citations counted in INSPIRE as of 17 Nov 2026


%\cite{Wu:2026pgc} 
\bibitem{Wu:2026pgc} H.~Wu and S.~Ying, ``From off-shell strings to entropy flows,'' [arXiv:2611.03827 [hep-th]]. %0 citations counted in INSPIRE as of 17 Nov 2026


%\cite{He:2014gva} 
\bibitem{He:2014gva} S.~He, T.~Numasawa, T.~Takayanagi and K.~Watanabe, ``Notes on Entanglement Entropy in String Theory,'' JHEP \textbf{05}, 106 (2015) doi:10.1007/JHEP05(2015)106 [arXiv:1412.5606 [hep-th]]. %65 citations counted in INSPIRE as of 16 Sep 2026


%\cite{Brustein:2022wiq} 
\bibitem{Brustein:2022wiq} R.~Brustein and Y.~Zigdon, ``On the entropy of strings and branes,'' JHEP \textbf{10}, 112 (2022) doi:10.1007/JHEP10(2022)112 [arXiv:2208.07372 [hep-th]]. %5 citations counted in INSPIRE as of 24 Aug 2026

%%Ahmadain Aron%%


%\cite{Ahmadain:2022tew} 
\bibitem{Ahmadain:2022tew} 
A.~Ahmadain and A.~C.~Wall, ``Off-shell strings I: S-matrix and action,'' SciPost Phys. \textbf{17}, no.1, 005 (2024) doi:10.21468/SciPostPhys.17.1.005 [arXiv:2211.08607 [hep-th]]. %24 citations counted in INSPIRE as of 11 Dec 2024


%\cite{Ahmadain:2022eso} 
\bibitem{Ahmadain:2022eso} 
A.~Ahmadain and A.~C.~Wall, ``Off-shell strings II: Black hole entropy,'' SciPost Phys. \textbf{17}, no.1, 006 (2024) doi:10.21468/SciPostPhys.17.1.006 [arXiv:2211.16448 [hep-th]]. %18 citations counted in INSPIRE as of 11 Dec 2024


%\cite{Halder:2023adw} 
\bibitem{Halder:2023adw} I.~Halder and D.~L.~Jafferis, ``Thermal Bekenstein-Hawking entropy from the worldsheet,'' JHEP \textbf{05}, 136 (2024) doi:10.1007/JHEP05(2024)136 [arXiv:2310.02313 [hep-th]]. %13 citations counted in INSPIRE as of 14 Aug 2026

%\cite{Mori:2025qoh} 
\bibitem{Mori:2025qoh} S.~Mori, T.~Sakai and M.~Shigemori, ``Black Hole Entropy from String Entanglement,'' [arXiv:2509.21796 [hep-th]]. %0 citations counted in INSPIRE as of 14 Aug 2026

%\cite{Ahmadain:2025pox} 
\bibitem{Ahmadain:2025pox} A.~Ahmadain and M.~Yang, ``Strings at the Tip of the Cone and Black Hole Entropy From the Worldsheet: Part I,'' [arXiv:2512.00637 [hep-th]]. %2 citations counted in INSPIRE as of 14 Aug 2026



%\cite{Fursaev:1995ef} 
\bibitem{Fursaev:1995ef} D.~V.~Fursaev and S.~N.~Solodukhin, ``On the description of the Riemannian geometry in the presence of conical defects,'' Phys. Rev. D \textbf{52}, 2133-2143 (1995) doi:10.1103/PhysRevD.52.2133 [arXiv:hep-th/9501127 [hep-th]]. %313 citations counted in INSPIRE as of 13 Oct 2026





%\cite{Brown:1986nw} 
\bibitem{Brown:1986nw} J.~D.~Brown and M.~Henneaux, ``Central Charges in the Canonical Realization of Asymptotic Symmetries: An Example from Three-Dimensional Gravity,'' Commun. Math. Phys. \textbf{104}, 207-226 (1986) doi:10.1007/BF01211590 %2922 citations counted in INSPIRE as of 17 Nov 2026




%\cite{Lewkowycz:2013nqa} 
\bibitem{Lewkowycz:2013nqa} A.~Lewkowycz and J.~Maldacena, ``Generalized gravitational entropy,'' JHEP \textbf{08}, 090 (2013) doi:10.1007/JHEP08(2013)090 [arXiv:1304.4926 [hep-th]]. %1342 citations counted in INSPIRE as of 13 Oct 2026


%\cite{Holzhey:1994we} 
\bibitem{Holzhey:1994we} C.~Holzhey, F.~Larsen and F.~Wilczek, ``Geometric and renormalized entropy in conformal field theory,'' Nucl. Phys. B \textbf{424}, 443-467 (1994) doi:10.1016/0550-3213(94)90402-2 [arXiv:hep-th/9403108 [hep-th]]. %1746 citations counted in INSPIRE as of 17 Nov 2026


%\cite{Calabrese:2004eu} 
\bibitem{Calabrese:2004eu} P.~Calabrese and J.~L.~Cardy, ``Entanglement entropy and quantum field theory,'' J. Stat. Mech. \textbf{0406}, P06002 (2004) doi:10.1088/1742-5468/2004/06/P06002 [arXiv:hep-th/0405152 [hep-th]]. %2939 citations counted in INSPIRE as of 17 Nov 2026


%BTZ

%\cite{Larsen:1994ah} 
\bibitem{Larsen:1994ah} A.~L.~Larsen and N.~G.~Sanchez, ``Strings propagating in the (2+1)-dimensional black hole anti-de Sitter space-time,'' Phys. Rev. D \textbf{50}, 7493-7518 (1994) doi:10.1103/PhysRevD.50.7493 [arXiv:hep-th/9405026 [hep-th]]. %54 citations counted in INSPIRE as of 13 Oct 2026

%\cite{Ashok:2021ffx} 
\bibitem{Ashok:2021ffx} S.~K.~Ashok and J.~Troost, ``Twisted strings in three-dimensional black holes,'' Eur. Phys. J. C \textbf{82}, no.10, 913 (2022) doi:10.1140/epjc/s10052-022-10903-y [arXiv:2112.08784 [hep-th]]. %15 citations counted in INSPIRE as of 13 Oct 2026




%\cite{Banados:1992wn} 
\bibitem{Banados:1992wn} M.~Banados, C.~Teitelboim and J.~Zanelli, ``The Black hole in three-dimensional space-time,'' Phys. Rev. Lett. \textbf{69}, 1849-1851 (1992) doi:10.1103/PhysRevLett.69.1849 [arXiv:hep-th/9204099 [hep-th]]. %3897 citations counted in INSPIRE as of 13 Oct 2026



%BTZ mixed flux

%\cite{Cagnazzo:2012se} 
\bibitem{Cagnazzo:2012se} A.~Cagnazzo and K.~Zarembo, ``B-field in AdS(3)/CFT(2) Correspondence and Integrability,'' JHEP \textbf{11}, 133 (2012) [erratum: JHEP \textbf{04}, 003 (2013)] doi:10.1007/JHEP11(2012)133 [arXiv:1209.4049 [hep-th]]. %176 citations counted in INSPIRE as of 13 Oct 2026


%\cite{Hoare:2013pma} 
\bibitem{Hoare:2013pma} B.~Hoare and A.~A.~Tseytlin, ``On string theory on $AdS_3 \times S^3 \times T^4$ with mixed 3-form flux: tree-level S-matrix,'' Nucl. Phys. B \textbf{873}, 682-727 (2013) doi:10.1016/j.nuclphysb.2013.05.005 [arXiv:1303.1037 [hep-th]]. %136 citations counted in INSPIRE as of 13 Oct 2026




%\cite{Maldacena:2000hw} 
\bibitem{Maldacena:2000hw} J.~M.~Maldacena and H.~Ooguri, ``Strings in AdS(3) and SL(2,R) WZW model 1.: The Spectrum,'' J. Math. Phys. \textbf{42}, 2929-2960 (2001) doi:10.1063/1.1377273 [arXiv:hep-th/0001053 [hep-th]]. %728 citations counted in INSPIRE as of 17 Nov 2026





%\cite{Horowitz:1993jc} 
\bibitem{Horowitz:1993jc} G.~T.~Horowitz and D.~L.~Welch, ``Exact three-dimensional black holes in string theory,'' Phys. Rev. Lett. \textbf{71}, 328-331 (1993) doi:10.1103/PhysRevLett.71.328 [arXiv:hep-th/9302126 [hep-th]]. %241 citations counted in INSPIRE as of 28 May 2025


%\cite{Polyakov:1981rd} 
\bibitem{Polyakov:1981rd} A.~M.~Polyakov, ``Quantum Geometry of Bosonic Strings,'' Phys. Lett. B \textbf{103}, 207-210 (1981) doi:10.1016/0370-2693(81)90743-7 %3334 citations counted in INSPIRE as of 17 Nov 2026


%\cite{Callan:1985ia} 
\bibitem{Callan:1985ia} C.~G.~Callan, Jr., E.~J.~Martinec, M.~J.~Perry and D.~Friedan, ``Strings in Background Fields,'' Nucl. Phys. B \textbf{262}, 593-609 (1985) doi:10.1016/0550-3213(85)90506-1 %2001 citations counted in INSPIRE as of 13 Oct 2026


%soliton

%\cite{Dabholkar:1990yf} 
\bibitem{Dabholkar:1990yf} A.~Dabholkar, G.~W.~Gibbons, J.~A.~Harvey and F.~Ruiz Ruiz, ``Superstrings and Solitons,'' Nucl. Phys. B \textbf{340}, 33-55 (1990) doi:10.1016/0550-3213(90)90157-9 %521 citations counted in INSPIRE as of 16 Aug 2025




%\cite{Casini:2011kv} 
\bibitem{Casini:2011kv} H.~Casini, M.~Huerta and R.~C.~Myers, ``Towards a derivation of holographic entanglement entropy,'' JHEP \textbf{05}, 036 (2011) doi:10.1007/JHEP05(2011)036 [arXiv:1102.0440 [hep-th]]. %1462 citations counted in INSPIRE as of 17 Nov 2026



%\cite{Zwiebach:2004tj} 
\bibitem{Zwiebach:2004tj} B.~Zwiebach, ``A first course in string theory,'' Cambridge University Press, 2006, ISBN 978-0-521-83143-7, 978-0-511-20757-0 %93 citations counted in INSPIRE as of 17 Nov 2026






%\cite{Maldacena:2001kr} 
\bibitem{Maldacena:2001kr} J.~M.~Maldacena, ``Eternal black holes in anti-de Sitter,'' JHEP \textbf{04}, 021 (2003) doi:10.1088/1126-6708/2003/04/021 [arXiv:hep-th/0106112 [hep-th]]. %1735 citations counted in INSPIRE as of 26 Oct 2026



%\cite{Affleck:1991tk} 
\bibitem{Affleck:1991tk} I.~Affleck and A.~W.~W.~Ludwig, ``Universal noninteger 'ground state degeneracy' in critical quantum systems,'' Phys. Rev. Lett. \textbf{67}, 161-164 (1991) doi:10.1103/PhysRevLett.67.161 %922 citations counted in INSPIRE as of 17 Nov 2026





%\cite{Takayanagi:2011zk} 
\bibitem{Takayanagi:2011zk} T.~Takayanagi, ``Holographic Dual of BCFT,'' Phys. Rev. Lett. \textbf{107}, 101602 (2011) doi:10.1103/PhysRevLett.107.101602 [arXiv:1105.5165 [hep-th]]. %644 citations counted in INSPIRE as of 17 Nov 2026


%\cite{Headrick:2010zt} 
\bibitem{Headrick:2010zt} M.~Headrick, ``Entanglement Renyi entropies in holographic theories,'' Phys. Rev. D \textbf{82}, 126010 (2010) doi:10.1103/PhysRevD.82.126010 [arXiv:1006.0047 [hep-th]]. %638 citations counted in INSPIRE as of 17 Nov 2026

















































 
 





\end{thebibliography}
\end{document}